\documentclass{emulateapj}
\usepackage{natbib}

\shorttitle{Nuclear Star Clusters in Edge-on Spirals}
\shortauthors{Seth, Dalcanton, Hodge, and Debattista}

\begin{document}

\slugcomment{Accepted for publication in the Astronomical Journal}

\title{Clues to Nuclear Star Cluster Formation from Edge-on Spirals}

\author{Anil C. Seth\footnote{Currently a CfA Postdoctoral Fellow, 60 Garden St., Cambridge, MA 02138, aseth@cfa.harvard.edu}}
\affil{University of Washington, Astronomy Dept., Box 351580, Seattle, WA 98195}

\author{Julianne J. Dalcanton\footnote{Alfred P. Sloan Research Fellow}}
\affil{University of Washington, Astronomy Dept., Box 351580, Seattle, WA 98195}

\author{Paul W. Hodge}
\affil{University of Washington, Astronomy Dept., Box 351580, Seattle, WA 98195}

\author{Victor P. Debattista\footnote{Brooks Prize Fellow}}
\affil{University of Washington, Astronomy Dept., Box 351580, Seattle, WA 98195 \\}

\begin{abstract}

We find 9 nuclear cluster candidates in a sample of 14 edge-on,
late-type galaxies observed with HST/ACS.  These clusters have
magnitudes ($M_I \sim -11$) and sizes ($r_{eff} \sim 3$~pc) similar to
those found in previous studies of face-on, late-type spirals and dE
galaxies.  However, three of the nuclear clusters are significantly
flattened and show evidence for multiple, coincident structural
components.  The elongations of these three clusters are aligned to
within $\sim$10$^\circ$ of the galaxies' major axes.  Structurally,
the flattened clusters are well fit by a combination of a spheroid and
a disk or ring, with the disk preferred in two of three cases.  The
nuclear cluster disks/rings have F606W-F814W ($\sim$V-I) colors
0.3-0.6 magnitudes bluer than the spheroid components, suggesting that
the stars in these components have ages $<$1~Gyr.  In NGC~4244, the
nearest of the nuclear clusters, we further constrain the stellar
populations via spectroscopy and multi-band photometry.  This nuclear
cluster is equally well fit by single-stellar populations with ages of
either $\sim$100~Myr or $\sim$1~Gyr, and with masses of
2-5$\times$10$^6$ M$_\odot$.  However, significantly better fits to
the spectroscopy and photometry are obtained by combining two or more
stellar populations.  Exploiting emission lines that appear to
originate $\sim$1\arcsec\ from the NGC~4244 nucleus, we determine a
lower limit on the dynamical mass of 2.5$^{+1.7}_{-1.2} \times
10^{6}$M$_\odot$ within 19~pc, typical of values found for other
nuclear clusters.  We also present tentative evidence that another of
the nuclear clusters (in NGC~4206) may also host a supermassive black
hole.  Based on our observational results we propose an {\it in situ}
formation mechanism for nuclear clusters in which stars form
episodically in compact nuclear disks, and then lose angular momentum
or heat vertically to form an older spheroidal structure.  We estimate
the period between star formation episodes to be $\sim$0.5 Gyr and
discuss possible mechanisms for tranforming the disk-like components
into spheroids.  We also note the connection between our objects and
massive globular clusters (e.g. $\omega$~Cen), ultra-compact dwarfs,
and supermassive black holes.

\end{abstract}
\keywords{galaxies:nuclei -- galaxies:star clusters -- galaxies:spiral -- galaxies:formation -- galaxies:individual (IC~5052, NGC~4244, NGC~4206)}

\section{Introduction}

Due to their deep gravitational potential wells and unique dynamics,
the centers of galaxies frequently host exotic objects such as
supermassive black holes, active galactic nuclei, and massive stellar
clusters.  Of these, only nuclear star clusters provide a visible
record of the accretion of stars and gas into the nucleus.  Therefore,
studies of nuclear star clusters hold the promise to
shed light on the formation of central massive objects of all types.

Bright nuclear star clusters have been observed in $\sim$75\% of the
77 face-on galaxies of type Scd and later studied by \citet{boker02}.
Similar clusters appear to be frequent in earlier type spirals as
well, although their study is complicated by the presence of bulges
and other nuclear phenomena \citep{carollo02}.  Dwarf elliptical
galaxies, which may originate from low-mass spiral galaxies
\citep[e.g.][]{mastropietro05,beasley06}, also frequently possess
nuclear star clusters \citep{grant05,cote06}.

In late-type spiral galaxies with little or no bulge, nuclear star
clusters are very prominent and can be studied in detail.  From their
sample of face-on, late-type spiral galaxies, \citet{boker04a} found
that nuclear clusters have dimensions similar to Milky Way globular
clusters, with typical half-light radii of 3.5 pc.  However, these
clusters are significantly brighter than globular clusters,
with $I$ band magnitudes ranging between $-8$ and $-16$
\citep{boker02}.  Their high luminosities are at least in part due to
their large masses; \citet{walcher05} have used high-resolution
spectra to derive typical masses $\sim$3$\times10^6$M$_\odot$ for nine
galaxies in the \citet{boker02} sample, similar in mass to the most
massive Milky Way cluster, $\omega$~Cen \citep{meylan86}.  However, the
brightness of nuclear clusters 
also results in part from the presence of young stellar populations.
Photometry and spectra of a number of clusters suggest that nuclear
clusters in both spiral and dwarf ellipticals have populations much
younger than globular clusters \citep[e.g.][]{ho95,lotz04}.  In
late-type galaxies, most clusters appear to have stars with ages
$<$100~Myr \citep{walcher06}.  
Furthermore, recent spectral studies
have shown that nuclear clusters are made up of composite stellar
populations, with most having substantial old ($\gtrsim$1~Gyr) stellar
component \citep{long02,sarzi05,walcher06,rossa06}.

Not only are nuclear clusters present in a large fraction of present
day spirals and dwarf ellipticals, they also have been invoked as the
progenitors of massive globular clusters and ultra-compact dwarfs.
In a hierarchical merging scenario, many galaxies will be tidally
stripped during mergers leaving behind only their dense nuclei
\citep[e.g.][]{bekki04}.  Nuclear cluster properties match the
half-light radii and exceptional luminosity of massive globulars and
UCDs \citep{mackey05,phillipps01,jones06}.  Furthermore, the multiple
stellar populations seen in $\omega$~Cen and G1
\citep[e.g.][]{bedin04,meylan01}, and the evidence for possible
self-enrichment among the brightest globular clusters in ellipticals
\citep[][]{harris06} strengthen the association between these bright
compact objects.

Finally nuclear clusters appear to be intimately connected to the
formation of supermassive black holes (SMBHs) at the centers of
galaxies.  For both spiral and elliptical galaxies, a clear
correlation is seen between nuclear cluster mass and bulge mass
\citep{cote06,wehner06,ferrarese06,rossa06}.  This correlation appears
to fall along the well-known $M_\bullet-\sigma$ relation between black
hole mass and bulge or galaxy mass
\citep[e.g.][]{ferrarese00b,ferrarese02}.  Both SMBHs and nuclear
clusters appear to contain the same fraction of the total galaxy mass,
supporting the idea that they are two related types of
central massive object \citep{wehner06}.  Whether a central massive
object is a nuclear cluster or a SMBH appears to depend on mass;
early-type galaxies more massive than $\sim$10$^{10}$~M$_\odot$ appear
to have SMBHs, while lower mass systems commonly host nuclear clusters
\citep{ferrarese06}.  These findings strongly suggest a similar
formation mechanism for nuclear clusters and SMBHs.  Increasing
evidence also points to massive star clusters as sites of
intermediate-mass black hole (IMBH) formation
\citep[e.g.][]{miller04,gebhardt05,patruno06}.  These IMBHs may also
follow the $M_\bullet-\sigma$ relation \citep{gebhardt02}.

In this context, a detailed understanding of how nuclear clusters
formed may reveal the general mechanism driving the formation of
central massive objects of all types.  Two basic scenarios have been
suggested \citep[e.g.][]{boker04a}: (1) nuclear clusters form from
multiple globular clusters accreted via dynamical friction
\citep[e.g.][]{tremaine75,lotz01}, and (2) nuclear clusters form {\it
in situ} from gas channeled into the center of galaxies
\citep{milosavljevic04}.  In this paper we present observational
evidence that favors the {\it in situ} formation scenario.  In
particular, we find nuclear clusters that are elongated along the
major axis of the galaxies and have two coincident components.
These clusters appear to consist of a
younger flattened disk or ring and an older spheroid.  
Based on the evidence we present, we suggest a detailed formation
mechanism in which nuclear clusters form in repeated star formation
events from very compact nuclear gas disks.

In \S\ref{obssec} we describe the imaging and spectroscopic
observations used in the paper.  \S\ref{morphsec} focuses on an
analysis of the morphology and luminosities of the clusters as
revealed by HST observations.  The stellar populations of the
multiple-component clusters are examined in \S\ref{popsec}.  In
\S\ref{masssec} we derive a dynamical mass for the NGC~4244 nuclear
cluster, and then in \S\ref{sbhsec} present tentative 
evidence that the NGC~4206 nucleus may harbor both a star cluster and
a supermassive black hole.  We present a detailed formation mechanism
and discuss it in the broader context in \S\ref{dissec}, and then
summarize our findings in \S\ref{sumsec}.

\section{Observations} \label{obssec}

\subsection{Identification of Nuclear Clusters}

Most of the data presented in this paper were taken with the Hubble
Space Telescope's (HST) Advanced Camera for Surveys (ACS) as part of a
Cycle 13 snapshot program.  Sixteen galaxies were observed in both the
F606W and F814W filters, with exposure times of $\sim$700 seconds.
Details of these observations and their reduction are presented in
\citet{seth05a} (Paper~I).  

We identified nuclear clusters (NCs) by careful examination of the
central regions of our galaxies.  The exact locations of the galaxy
centers were determined using fits to 2MASS data presented in Paper~I.
Nuclear clusters were distinguished based on their proximity to the
photocenter of the galaxy, their brightness/prominence, and their
extent (i.e. if they appeared resolved).  At the distance of most of
our galaxies, NCs such as those observed by
\citet{boker02} should be at least partially resolved.

Table~\ref{sampletab} lists the galaxies in our sample and whether
each possesses a NC (shown in the fifth column).  In two of the
sixteen galaxies in the sample, the central region was not covered by
the ACS images (denoted with a ``N/A'' in Table~\ref{sampletab}).
Of the remaining 14 galaxies, seven have prominent and/or extended
sources located within $\sim$2 arcseconds of the photometric center of
the galaxy.  Another two have possible detections of NCs, where the
candidate clusters were not very prominent (denoted with a ``Maybe'').
The remaining five galaxies have no obvious candidate NCs.  Of these
five definite non-detections, three (IRAS~06070-6147, IRAS~07568-4942,
\& NGC~4631) are in galaxies where strong dust lanes could have
easily obscured any extant cluster.  The other two of the
non-detections (NGC~55 and IC~2233) are in low mass galaxies without
strong dust lanes, suggesting that these galaxies truly lack NCs.  

\begin{figure*}
\plotone{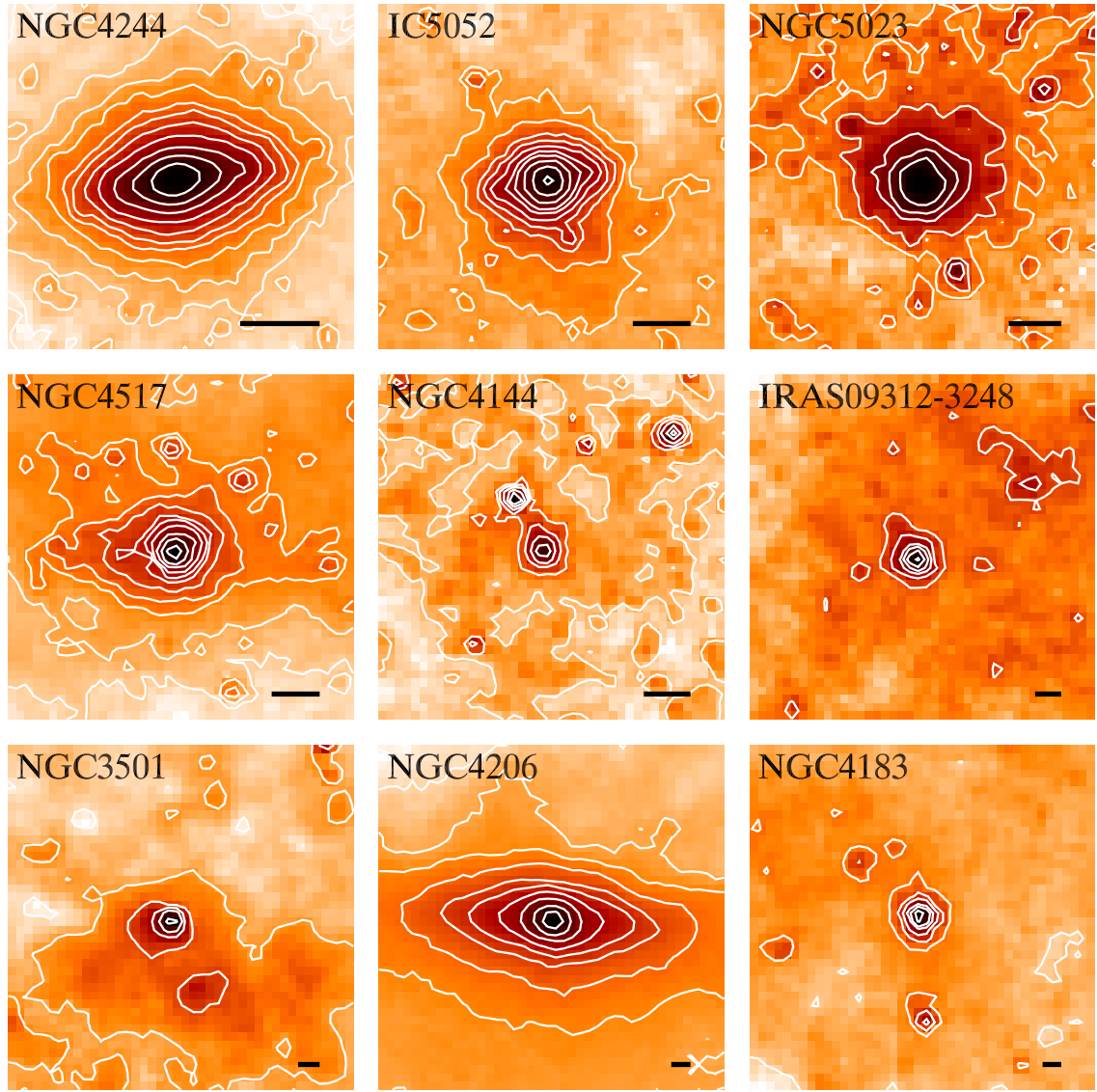}
\caption{Images in the F814W filter of the nine candidate nuclear
clusters.  Each image is 2\arcsec\ on a side, and the bar in the bottom
right corner indicates a length of 10 pc.  Images have been rotated so
that the x-axis lies along the position angle of the major axis of the
galaxy disk.}
 \label{allclusterfig}
\end{figure*}

\begin{deluxetable*}{lcccccccc}
\tablecaption{Galaxy Sample and Nuclear Cluster Properties\label{sampletab}}
\tablewidth{0pt}
\tabletypesize{\scriptsize}
\tablehead{
     \colhead{Galaxy}  &
     \colhead{Type}  &
     \colhead{$M_B$}  &
     \colhead{$V_{max}$}  &
     \colhead{Possess} &
     \colhead{Multiple} &
     \colhead{$M_{F814W}$} &
     \colhead{$m_{F606W}-m_{F814W}$} &
     \colhead{m-M} \\
     \colhead{}  &
     \colhead{}  &
     \colhead{}  &
     \colhead{[km/sec]}  &
     \colhead{NC} &
     \colhead{Components} &
     \colhead{} &
     \colhead{} &
     \colhead{}
}

\startdata
IC~2233         & SBcd  & -16.60 & 79   & No       & N/A  &        &      & 29.96\tablenotemark{b} \\
IC~5052         & SBcd  & -17.23 & 79   & Yes      & Yes  & -10.77 & 1.08 & 28.90\tablenotemark{a} \\ 
IRAS~06070-6147 & Sc    & -20.50 & 133  & No;Dust  & N/A  & 	  &      & 31.22\tablenotemark{d} \\ 
IRAS~07568-4942 & SBc   & -17.70 & 142  & No;Dust  & N/A  &        &      & 30.06\tablenotemark{c} \\ 
IRAS~09312-3248 & Sc    & -17.90 & 98   & Maybe    & No   &  -9.24 & 1.36 & 30.61\tablenotemark{d} \\ 
NGC~55          & SBm   & -17.05 & 67   & No       & N/A  &        &      & 26.63\tablenotemark{a} \\ 
NGC~891         & Sb    & -18.78 & 214  & N/A      & N/A  &        &      & 29.61\tablenotemark{e} \\ 
NGC~3501        & Sc    & -17.37 & 136  & Yes      & No   & -10.58 & 1.62 & 31.04\tablenotemark{d} \\ 
NGC~4144        & SBc   & -17.20 & 67   & Maybe    & No   &  -8.33 & 0.85 & 29.36\tablenotemark{a} \\ 
NGC~4183        & Sc    & -18.41 & 103  & Yes      & No   & -11.15 & 0.93 & 31.35\tablenotemark{f} \\ 
NGC~4206        & Sbc   & -18.46 & 124  & Yes      & Yes  & -14.17 & 0.93 & 31.29\tablenotemark{g} \\ 
NGC~4244        & Sc    & -17.08 & 93   & Yes      & Yes  & $<$-12.65\tablenotemark{i}& 0.77	& 28.20\tablenotemark{a} \\ 
NGC~4517        & Sc    & -18.18 & 137  & Yes      & Maybe& -11.34 & 1.77 & 29.30\tablenotemark{c} \\ 
NGC~4565        & Sb    & -20.47 & 227  & N/A      & N/A  &	 &       & 31.05\tablenotemark{h} \\ 
NGC~4631        & SBcd  & -19.75 & 131  & No;Dust  & N/A  &	 &       & 29.42\tablenotemark{a} \\ 
NGC~5023        & Sc    & -16.29 & 77   & Yes      & Maybe&  -9.79 & 1.26 & 29.10\tablenotemark{a} \\ 
\enddata
\tablecomments{ Types and velocities from HYPERLEDA/LEDA
  \citep{paturel95,paturel03}.  (a) \citet{seth05a} (b)
  \citet{bottinelli85}, (c) \citet{bottinelli88}, (d) assuming
  H$_0$=70 km/sec, (e) \citet{tonry01}, (f) \citet{tully00}, (g)
  \citet{gavazzi99}, (h) \citet{jensen03} (i) The derived magnitudes on the NGC~4244 cluster are lower limits due to the central pixels being saturated in both bands.  }

\end{deluxetable*}

In summary, we have detected candidate NCs in 50\% to 65\% of our
sample.  Eliminating the galaxies with strong dust lanes, the fraction
of galaxies with detected NCs may be as high as 82\%.  These detection
fractions are consistent with other studies, both for late-type
galaxies, where \citet{boker02} finds $\sim$75\% of systems have
clusters, and for early-type galaxies observed as part of the ACS
Virgo Cluster Survey \citep{cote06}.  However, our sample is less
well-suited than these studies for constraining the frequency of
nuclear clusters because the edge-on perspective can result in more
chance superpositions and greater obscuration from dust.

\subsection{Apache Point Observatory 3.5-m Spectrum}

To better constrain the stellar populations of the NGC~4244 nuclear
cluster (the nearest in our sample), we obtained long-slit spectra on
May 19, 2005 with the Double Imaging Spectrograph (DIS) on the Apache
Point Observatory (APO) 3.5m telescope.  The DIS spectrograph obtains
both blue and red spectra simultaneously.  We used the
``medium-resolution'' blue grating (1.22\AA/pixel) and the
``high-resolution'' red grating (0.84\AA/pixel).  The resulting images
had good quality spectral coverage from 3700\AA~to 5500\AA~on the blue
side and 5600\AA~to 7200\AA~on the red side.  A 1.5\arcsec\ slit was
used.
Two exposures of 15 and 30 minutes were obtained centered on the
nuclear cluster and oriented along the major axis of NGC~4244 with a
position angle of -43$^\circ$.  We note that the elongation of the
nuclear cluster is aligned to within $\sim$10$^\circ$ of the galaxy's
major axis (see Fig.~\ref{allclusterfig}).  Seeing at the time of
observation was $\sim$1.0\arcsec, and thus the cluster was essentially
unresolved.  We reduced these spectra using a PyRAF pipeline designed
for DIS data reduction (written by Kevin Covey) which utilizes
standard IRAF {\sc onedspec} and {\sc twodspec} routines.  The
NGC~4244 cluster observations were wavelength calibrated using lamp
spectra, and a further correction was applied to remove small shifts
visible in the sky lines.  The velocities were calibrated to the local
standard of rest using {\sc rvcorrect} and {\sc dopcor}. Observations
of Feige~66 at nearly identical airmass were obtained for flux
calibration shortly before the observations; however the night was not
photometric and thus the absolute flux scale is not fully constrained.
The two observations of the NGC~4244 cluster differ in flux scale by
under 10\%, and our absolute flux is probably uncertain by a similar
amount.  Combined, the two spectra give a S/N of $\sim$35 per pixel.
The cluster spectrum was extracted from within a 2\arcsec\ diameter aperture,
and regions used to define the background subtraction were within the
confines of the galaxy, 20\arcsec-30\arcsec\ ($\sim$1/4th the scale
length of the galaxy) along the major axis in both directions.  Any
contamination from the underlying galaxy component should be very
small, as the source was well above the background at all wavelengths.
We analyze this data in \S4 and \S5.

\section{Nuclear Cluster Morphologies and Luminosities} \label{morphsec}

The most basic result of this paper is that nuclear clusters are not
all simple, single-component objects, as is evident from inspection
of Figure~\ref{allclusterfig}, which shows the F814W images of all
nine nuclear cluster candidates.  Of these candidates three (IC~5052,
NGC~4206, \& NGC~4244) look like miniature S0 galaxies,
possessing both an elongated disk or ring component and a spheroidal
component.  These components are both compact with physical dimensions
of tens of parsecs or less.  We will show in \S\ref{morphsec}.1.3 that
these components appear to physically overlap.  We therefore refer to
these clusters as multi-component clusters throughout the rest of the
paper.
Two other galaxies, NGC~4517 \& NGC~5023 also show tentative evidence
for similar structures.  In this section we focus on modeling the
morphology of the clusters, in particular those that have multiple
components.  We will show: (1) the nuclear cluster elongations are
closely aligned with the disk of the host galaxy, (2) models with both
a spheroidal and disk or ring components fit the multi-component
clusters significantly better than single-component models, and (3)
the nuclear clusters in our sample have magnitudes and sizes similar
to those observed in other galaxies.

\subsection{Model fitting} 

The dimensions and scales of the clusters were determined via 2-D
fitting to analytical functions (e.g. King profiles).  Because the
objects are very compact, it was necessary to include the effects of
the PSF.  We used two similar methods to fit the nuclear clusters.
First, we used the {\tt ishape} program \citep{larsen99},
which was designed to fit semi-resolved clusters.  Unfortunately, {\tt
ishape} only works with elliptically symmetric objects,
which some of our objects clearly are not.  Therefore, we wrote
our own code in IDL, using the existing {\tt MPFIT2DFUN} package to
perform Levenberg-Marquardt least squares fits.  

In both {\tt ishape} and our program, analytical functions are
convolved with the PSF before fitting to the data.  We performed this
convolution using a position-variable PSF derived from our data (see
Paper~I for details), and over-sampled by a factor of 10 to minimize 
the effects of aliasing and binning.  A separate PSF was generated for
each cluster based on its position on the chip.  We conducted tests to
insure that our program give similar results to the {\tt ishape}
program.

We used {\tt ishape} (1) to determine what types of functions best fit
the data, and (2) to assess if each cluster is indeed resolved.  We
fit Gaussian, King, Moffat, and Hubble profiles to each cluster and
found that King and Moffat profiles provided the best overall fits in
terms of the reduced $\chi^2$,  as also found by \citet{boker04a}.
Both profiles provided equally good fits to our clusters, and we chose
to continue our fitting with elliptical King profiles of the form:

\begin{eqnarray}
\Sigma(z)  & = & \Sigma_{0} \left(\frac{1}{\sqrt{1+(\frac{z}{r_{core}})^2}} - 
 \frac{1}{\sqrt{1+(\frac{r_{tidal}}{r_{core}})^2}}\right) \\ \nonumber
& & {\rm with} \ z=\sqrt{x^2+\left(\frac{z}{q}\right)^2}
\end{eqnarray}
where $\Sigma(z)$ is the projected surface brightness, $x$ and $z$ are
the coordinates along the major and minor axes, and $q$ is the axial
ratio (minor/major).  We have fixed the concentration ($c \equiv
r_{tidal}/r_{core}$) to be 15 for our fits, based on the fit
concentrations of our brightest clusters and similar to that found for
Galactic analog $\omega$~Cen \citep{trager95}.  We report only
half-light/effective radii as this quantity is well determined even
for relatively low signal-to-noise data \citep{carlson01}, and does
not depend strongly on the chosen model \citep[see
Fig.~2,][]{boker04a}.

Given sufficient signal-to-noise, semi-extended sources can be
detected with intrinsic widths as small as 10\% of the PSF
\citep{larsen99}. 
The FWHM of our PSF is $\sim$0.9 pixels, meaning objects with widths
smaller than a pixel should be easily resolvable.  The signal-to-noise
of our fits ranged from $\sim$50-1500.  For four of the galaxies
(IC~5052, NGC~4206, NGC~4244, \& NGC~5023), the FWHM of the fitted
{\tt ishape} King profiles is greater than one pixel, and the objects
are clearly partially resolved.  The $\chi^2$ values of these
fits is 2-20 times better than the $\chi^2$ of a point-source fit (see
last column, Table~\ref{profiletab}).  The remaining fits have FWHM
$>$0.2 pixels, and have $\chi^2$ values 1-3$\times$ the best-fitting
point-source $\chi^2$.  Of these, NGC~4144 and NGC~4517 are still
clearly resolved, while IRAS~09312-3248, NGC~3501, \& NGC~4183 (all of
which are at large distances) are only marginally resolved.

\begin{deluxetable*}{lccccccc} 
\tablewidth{0pt}
\tabletypesize{\scriptsize}
\tablecaption{Elliptical King Profile Fits\label{profiletab}}
\tablehead{
     \colhead{Galaxy}  &
     \colhead{Pix. Scale} &
     \colhead{Filter} &
     \colhead{$r_{eff}$} &
     \colhead{$q$} &
     \colhead{$\Delta$PA\tablenotemark{a}} &
     \colhead{Reduced} &
     \colhead{$\chi^2_{PS}/\chi^2$} \\
     \colhead{}  &
     \colhead{[pc/pixel]} &
     \colhead{} &
     \colhead{[pc]} &
     \colhead{($b/a$)} &
     \colhead{[$\circ$]} &
     \colhead{$\chi^2$} &
     \colhead{[{\tt ishape}]}
} 
\startdata
        IC~5052  &    1.46  &   F606W  &    3.64  &    0.76  &    -3.8  &  23.8 &   5.78 \\
        IC~5052  &    1.46  &   F814W  &    2.93  &    0.80  &    -6.6  &  18.1 &   7.00 \\
IRAS~09312-3248  &    3.21  &   F606W  &    2.46  &    0.68  &    43.1  &   2.8 &   1.50 \\
IRAS~09312-3248  &    3.21  &   F814W  &    2.24  &    0.86  &    39.8  &   3.5 &   1.42 \\
       NGC~3501  &    3.91  &   F606W  &    6.82  &    1.00  &    89.9  &   8.2 &   1.41 \\
       NGC~3501  &    3.91  &   F814W  &    3.58  &    0.61  &    19.2  &  10.1 &   1.38 \\
       NGC~4144  &    1.81  &   F606W  &    1.46  &    0.85  &    83.4  &   3.2 &   2.69 \\
       NGC~4144  &    1.81  &   F814W  &    1.99  &    0.81  &    74.8  &   4.6 &   2.30 \\
       NGC~4183  &    4.51  &   F606W  &    2.76  &    1.00  &   -40.8  &   9.6 &   1.20 \\
       NGC~4183  &    4.51  &   F814W  &    3.02  &    0.78  &    68.8  &   6.4 &   1.12 \\
       NGC~4206  &    4.39  &   F606W  &   16.30  &    0.37  &     5.6  &  54.7 &   2.39 \\
       NGC~4206  &    4.39  &   F814W  &   15.81  &    0.40  &     5.3  &  42.4 &   2.75 \\
       NGC~4244  &    1.06  &   F606W  &    2.92  &    0.46  &   -10.0  &  94.1 &  20.73 \\
       NGC~4244  &    1.06  &   F814W  &    3.35  &    0.49  &    -9.9  &  82.3 &  19.63 \\
       NGC~4517  &    1.76  &   F606W  &    1.71  &    0.85  &   -22.9  &  46.0 &   1.20 \\
       NGC~4517  &    1.76  &   F814W  &    1.33  &    0.85  &   -16.1  &  64.2 &   1.23 \\
       NGC~5023  &    1.60  &   F606W  &   12.54  &    0.81  &    16.2  &  13.4 &   4.27 \\
       NGC~5023  &    1.60  &   F814W  &   10.22  &    0.83  &    18.6  &  11.1 &   6.89 \\

\enddata
\tablecomments{{\it (a) } $\Delta$PA is the position angle relative to the galaxy disk.}

\end{deluxetable*}

\subsubsection{Elliptical King Model Fits} 

Table~\ref{profiletab} shows the results of our program's fits to
single-component 
elliptical King profiles with $c=15$.  These fits show:
\begin{enumerate}

\item All the well resolved clusters are flattened, as shown in
Figure~\ref{qfig}.  The median axis ratio ($q \equiv b/a$) is 0.81,
with $q \sim 0.4$ for NGC~4206 and NGC~4244, the two systems with the
most prominent disk (see \S\ref{morphsec}.1.2). 

\begin{figure}
\plotone{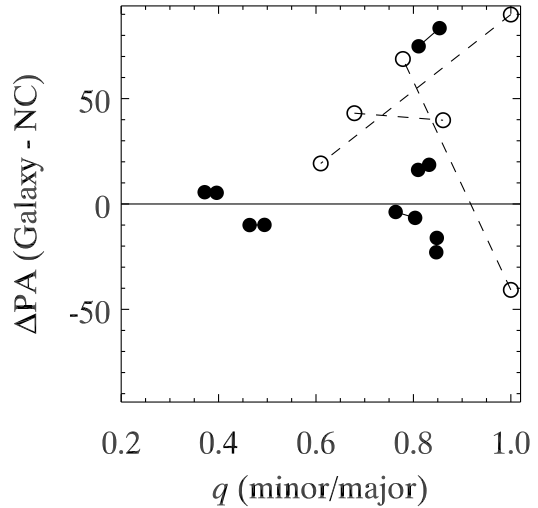}
\caption{Elliptical King profile fit results: the axial ratio $q$ is
plotted against the position angle of the major axis of the nuclear
cluster relative to the galaxy disk ($\Delta$PA).  The connected data
points connect the fits for each galaxy in the two filters.  Open
circles indicate the three galaxies (IRAS09312-3248, NGC~3501,
NGC~4183) for which good fits were not achieved, due to the NC
compactness and complexity of the surrounding emission.  This figure
shows that most of the clusters we have detected were significantly
flattened, and that the flattening aligns closely with the plane of
the galaxy.\\ }
\label{qfig}
\end{figure}

\item The position angle of the three multi-component nuclear clusters
(IC~5052, NGC~4206, \& NGC~4244) are all aligned within 10 degrees of the
galaxy disk position angle.  This can be seen plainly in
Figures~\ref{allclusterfig} and \ref{qfig}.  The NGC~4517 and NGC~5023
clusters also are somewhat elongated along the plane of the galaxy.

\item The effective radius (also known as the half-light radius)
ranges between 1 and 20 pc, with most of the clusters having effective
radii between 1-4 pc.  Figure~\ref{magfig} demonstrates that the
measured effective radii are similar to those observed for other objects
including late-type nuclear clusters \citep[filled
circles][]{boker04a} and galactic globular clusters \citep[open
triangles][]{harris96}.

\item After subtraction of the King profile fit, the residuals of
IC~5052, NGC~4206, and NGC~4244 clearly show evidence for a flat
disk- or ring-like component (see the middle column of Fig.~\ref{residfig}).
The reduced $\chi^2$ of these three galaxies is also very high,
suggesting that the single component King profile is not a good fit to
the data.  We note that NGC~4517, which has a disk-like appearance in
its outer isophotes, also has a high $\chi^2$ value.  In NGC~4244 the
central eight pixels were saturated and therefore excluded from the
fit.

\end{enumerate}

\begin{figure*}
\epsscale{0.7}
\plotone{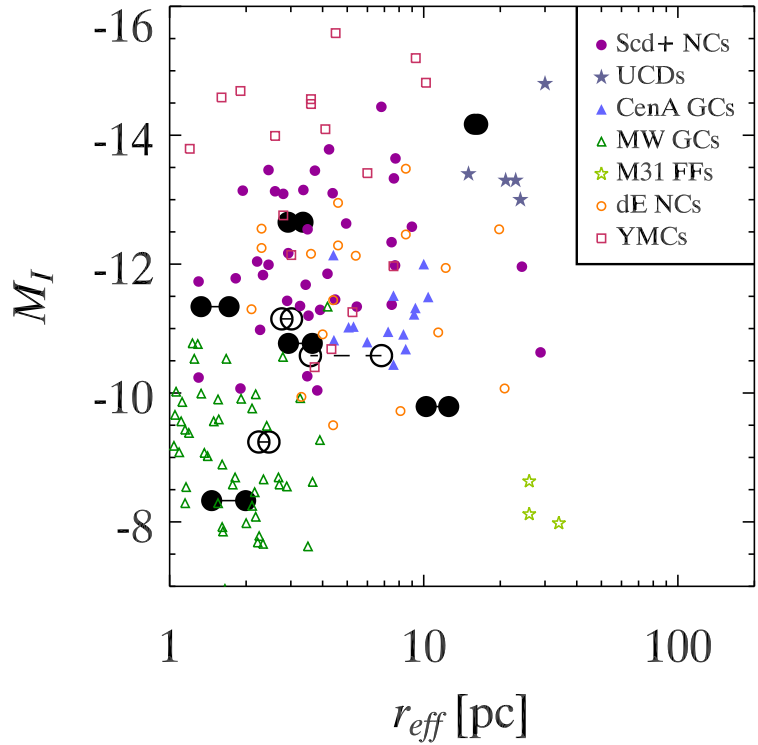}
\epsscale{1.0}
\caption{The fitted effective radius vs. the absolute aperture
magnitude in the F814W filter.  Large filled and open circles as in
Fig.~\ref{qfig}.  Small symbols denote the following: {\it Filled
circles -- } late-type nuclear clusters from \citet{boker04a}. {\it
Open circles -- } dE,N nuclei from \citet{depropris05} (transformed to
$M_I$ using Padova models which suggests stellar populations older
than 1~Gyr have F814W-F850LP colors of 0.25$\pm$0.1 mags). {\it Filled
triangles -- } massive globular clusters in NGC~5128 observed by
\citet{martini04}.  {\it Open triangles -- } Galactic globular
clusters from the \citet{harris96} catalog (note that $\omega$~Cen is
the brightest of these).  {\it Filled Stars -- } Fornax UCDs from
\citet{drinkwater03}, assuming V-I of 1.1$\pm$0.1 as shown in
\citet{mieske02}.  {\it Open Stars -- } faint fuzzy M31 clusters from
\citet{huxor05}.  {\it Open Squares -- } young massive clusters (YMCs)
from \citet{bastian06}.  Note that Padova models suggest $M_{F814W}
\sim M_I$ to within a few hundreths of a magnitude.  }
\label{magfig}
\end{figure*}

\begin{figure}
\plotone{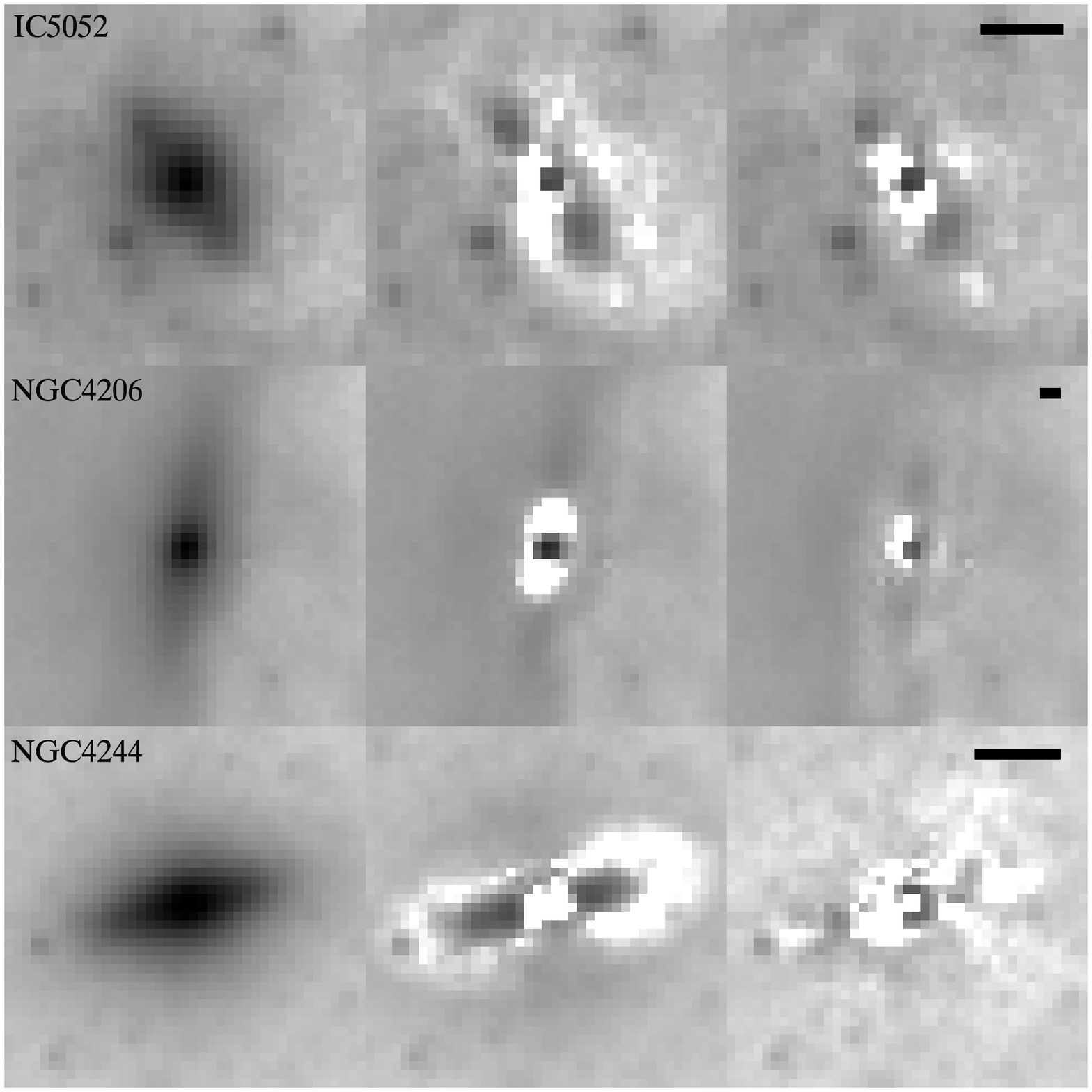}
\caption{Residuals of model fits to F606W observations of nuclear
clusters in IC~5052, NGC~4206, and NGC~4244.  The images show the area
over which fits were done, and are in the coordinate system of the ACS
images.  They are thus rotated from the images in
Figs.~\ref{allclusterfig} and \ref{colorfig}.  Bars in the upper right
show a length of 10~pc in each galaxy.  The left most column shows the
original image, the middle and right columns show the images after
subtraction of the best-fitting elliptical king and elliptical King
$+$ disk profiles.  All are shown with the same scaling.  In each
case, the elliptical King $+$ disk profile clearly provides a better
fit to the cluster than a single-component elliptical King profile. \\}
 \label{residfig}
\end{figure}

\subsubsection{Elliptical King $+$ Disk Models}

To account for the flattened residuals remaining after fits to the
elliptical king profiles in the IC~5052, NGC~4206, and NGC~4244
nuclear clusters, we have fit an additional morphological component to
these clusters.  In this section we examine fits to disk models, in
the next section we consider fits to ring models and compare the disk
and ring fits.

As a disk model we adopted the surface brightness profile of a
transparent, self-gravitating, edge-on exponential disk
\citep{vanderkruit81}:
\begin{equation}
\Sigma(x,z) = \Sigma_{0} \frac{x}{h_x} K_{1}\left(\frac{x}{h_x}\right) sech^{2}\left(\frac{z}{z_{0}}\right)
\end{equation}
where $x$ and $z$ are the major and minor axis coordinates, $\Sigma_0$
is the central surface brightness, $K_1$ is the modified Bessel
function, $h_x$ is the disk's exponential scale length and $z_0$ is
the scale height of the disk.  We note that we tried other vertical
profile fits (e.g. simple exponential) which were too concentrated
near the midplane to provide good fits to the data.

Figure~\ref{residfig} shows that the inclusion of a disk component
clearly improves the fit relative to the single component elliptical
King profile.  Quantitatively, the $\chi^2$ of the fits shown in
Table~\ref{disktab} is significantly reduced relative to those for the
single-component fit (Table~\ref{profiletab}).  The improvement is
most dramatic in NGC~4244, and is least significant in IC~5052.
However, the reduced $\chi^2$ values remain well above unity
suggesting the model doesn't fit the data perfectly.  For IC~5052 and
NGC~4244, the disk scale lengths are comparable to the effective radii
of the spheroidal components ($h_x \sim 3$~pc), while in NGC~4206 the
disk component is very extended ($h_x = 31$~pc).  

In NGC~4206 and IC~5052, the residuals show a similar pattern which
suggests the presence of a compact central source and/or a ring rather
than a disk.  We discuss the issue of whether the components are rings
or disks in the section below.  Both these galaxies also show a
color-gradient perpendicular to the major axis, suggesting the
presence of a thin dust disk.

\subsubsection{Rings vs. Disks}

In this section we present evidence that the flattened components in
NGC~4206 and NGC~4244 clusters are
better fit by a disk model than a ring model, while in IC~5052, a ring
model provides a better fit.  In all three cases, it appears that the
disk or ring extends to small radii and overlaps with the light
distribution of the spheroidal component.

In Table~\ref{ringtab} we present the best-fitting ring $+$ elliptical
king models for each image.  For the ring component, we assumed a
constant midplane density with an outer ($r_{out}$) and inner
($r_{in}$) radius:
\begin{eqnarray}
& & {\rm if}\ x < r_{in}:\quad  \\ \nonumber
& & \; \; \Sigma(x,z)  = \,2 \rho_{0} \, sech^{2}\left(\frac{z}{z_{0}}\right)
\left(\sqrt{r_{out}^2 -  x^2} - \sqrt{r_{in}^2- x^2}\right)   \\ \nonumber
& & {\rm if}\ r_{in} < x < r_{out}:\quad  \\ \nonumber
& & \; \; \Sigma(x,z)  =\,2 \rho_{0} \, sech^{2}\left(\frac{z}{z_{0}}\right)
\left(\sqrt{r_{out}^2 -  x^2}\right)  \\ \nonumber
& & {\rm else}:\quad  \Sigma(x,z)  = \,0
\end{eqnarray}
This model is brightest for $r_{in} < x < r_{out}$, and
contributes less flux near the center, making it qualitatively
different from the disk model (Eq.~2).  However, given that the
details of both models are somewhat arbitrary, the results of the fits
should be viewed with some caution. 

\begin{deluxetable*}{lcc|ccc|ccc|cc} 
\tablewidth{0pt}
\tabletypesize{\scriptsize}
\tablecaption{Elliptical King $+$ Disk Profile Fits\label{disktab}}
\tablehead{
     \colhead{Galaxy}  &
     \colhead{Pix. Scale} &
     \colhead{Filter} &
     \multicolumn{3}{c}{Disk} &
     \multicolumn{3}{c}{Ell. King} &
     \colhead{$\Delta$PA} &
     \colhead{Reduced} \\
     \colhead{}  &
     \colhead{[pc/pixel]} &
     \colhead{} &
     \colhead{$\mu_0$} &
     \colhead{$h_{x}$} &
     \colhead{$z_0$} &
     \colhead{$\mu_0$} &
     \colhead{$r_{eff}$} &
     \colhead{$q$} &
     \colhead{[$\circ$]} &
     \colhead{$\chi^2$} \\
     \colhead{}  &
     \colhead{} &
     \colhead{} &
     \colhead{[mag/$\Box$\arcsec]} &
     \colhead{[pc]} &
     \colhead{[pc]} &
     \colhead{[mag/$\Box$\arcsec]} &
     \colhead{[pc]} &
     \colhead{} &
     \colhead{} &
     \colhead{}
} 
\startdata
  IC~5052  &    1.46  &   F606W  &   17.27  &    3.54  &    1.47  &   16.15  &    3.97  &    1.00  &   -10.8  &    20.3 \\
  IC~5052  &    1.46  &   F814W  &   16.28  &    3.48  &    1.19  &   14.74  &    3.55  &    1.00  &   -12.0  &    15.7 \\
 NGC~4206  &    4.39  &   F606W  &   17.44  &   31.00  &   15.76  &   12.87  &    4.94  &    0.45  &     5.1  &    26.8 \\
 NGC~4206  &    4.39  &   F814W  &   16.61  &   31.36  &   17.42  &   12.00  &    5.32  &    0.44  &     5.0  &    16.7 \\
 NGC~4244  &    1.06  &   F606W  &   13.64  &    2.51  &    1.64  &   16.29  &   10.79  &    0.78  &   -10.1  &    15.1 \\
 NGC~4244  &    1.06  &   F814W  &   13.18  &    2.73  &    1.42  &   13.84  &    5.70  &    0.73  &    -9.6  &    16.3

\enddata

\end{deluxetable*}

\begin{deluxetable*}{lc|cccc|ccc|cc} 
\tablewidth{0pt}
\tabletypesize{\scriptsize}
\tablecaption{Elliptical King $+$ Ring Profile Fits\label{ringtab}}
\tablehead{
     \colhead{Galaxy}  &
     \colhead{Filter} &
     \multicolumn{4}{c}{Ring} &
     \multicolumn{3}{c}{Ell. King} &
     \colhead{$\Delta$PA} &
     \colhead{Reduced} \\
     \colhead{}  &
     \colhead{} &
     \colhead{$r_{out}$} &
     \colhead{$r_{in}$} &
     \colhead{$z_0$} &
     \colhead{$L_{ring}/$} &
     \colhead{$\mu_0$} &
     \colhead{$r_{eff}$} &
     \colhead{$q$} &
     \colhead{[$\circ$]} &
     \colhead{$\chi^2$} \\
     \colhead{}  &
     \colhead{} &
     \colhead{[pc]} &
     \colhead{[pc]} &
     \colhead{[pc]} &
     \colhead{$L_{tot}$} &
     \colhead{[mag/$\Box$\arcsec]} &
     \colhead{[pc]} &
     \colhead{} &
     \colhead{}
} 
\startdata
  IC~5052  &   F606W  &   11.83  &    6.83  &    1.44  &    0.18  &   15.64  &    3.24  &    1.00  &   -11.0  &    16.9 \\
  IC~5052  &   F814W  &   12.13  &    6.91  &    1.61  &    0.14  &   14.36  &    3.05  &    1.00  &   -11.5  &    13.2 \\
 NGC~4206  &   F606W  &   84.82  &    0.44  &   18.56  &    0.53  &   13.57  &    7.69  &    0.48  &     5.3  &    32.5 \\
 NGC~4206  &   F814W  &   83.78  &    0.44  &   20.69  &    0.53  &   12.59  &    7.70  &    0.47  &     5.2  &    21.5 \\
 NGC~4244  &   F606W  &   10.42  &    0.11  &    0.92  &    0.18  &   13.19  &    4.07  &    0.58  &   -10.1  &    26.7 \\
 NGC~4244  &   F814W  &   10.75  &    0.14  &    0.91  &    0.16  &   12.51  &    4.25  &    0.59  &    -9.7  &    23.5 
\enddata
\end{deluxetable*}

For the NGC~4206 and NGC~4244 NCs the ring fits (Table~\ref{ringtab})
have reduced $\chi^2$ values that are 20-75\% higher than the disk
values shown in Table~\ref{disktab}.  Furthermore, the best-fitting
ring models have inner radii extending all the way to the center of
cluster. Thus, the evidence suggests that in the NGC~4206 and NGC~4244
NCs, the second, flattened component is disk-like rather than being
ring-like.

\begin{figure*}
\epsscale{0.4}
\plotone{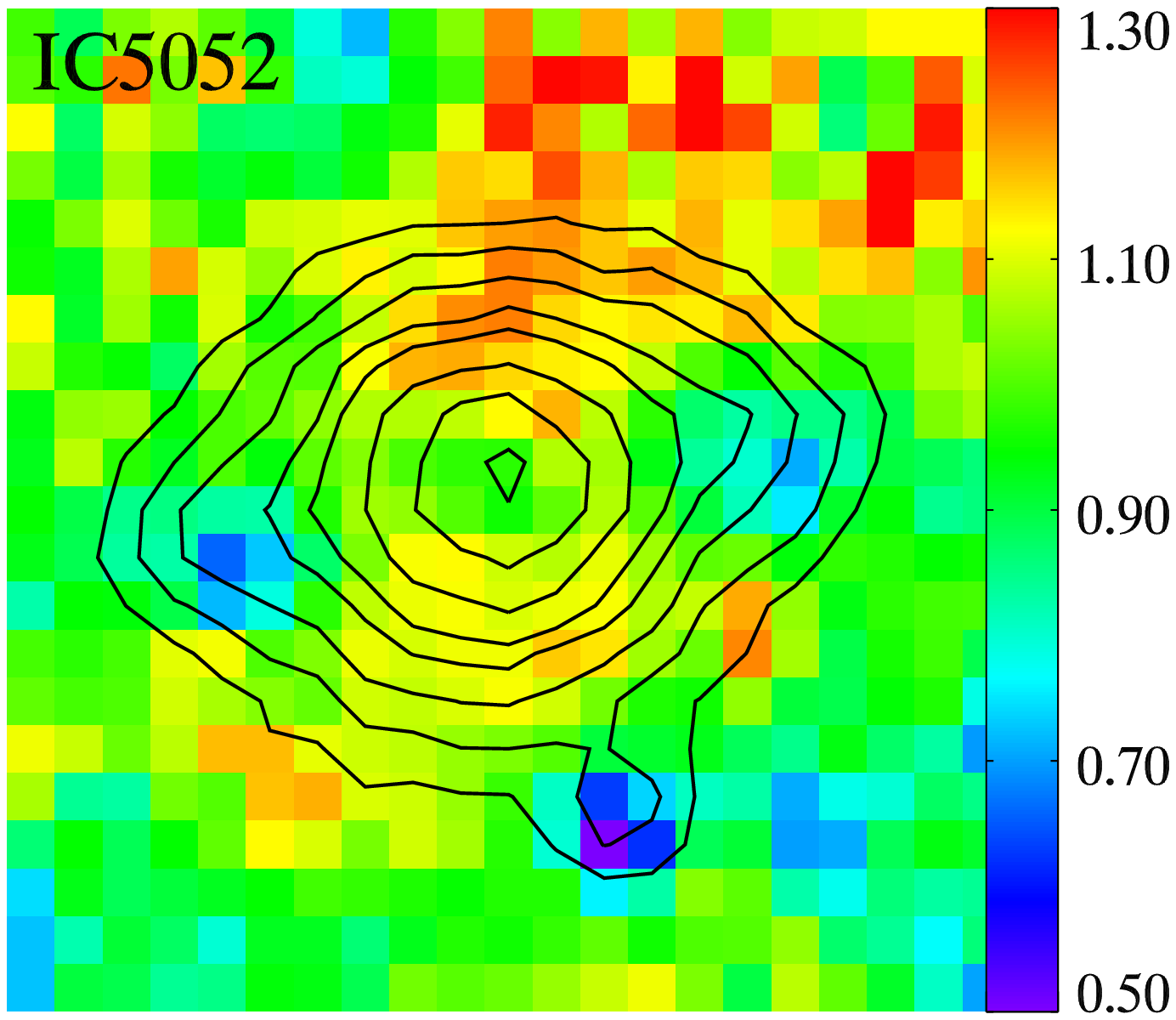}
\plotone{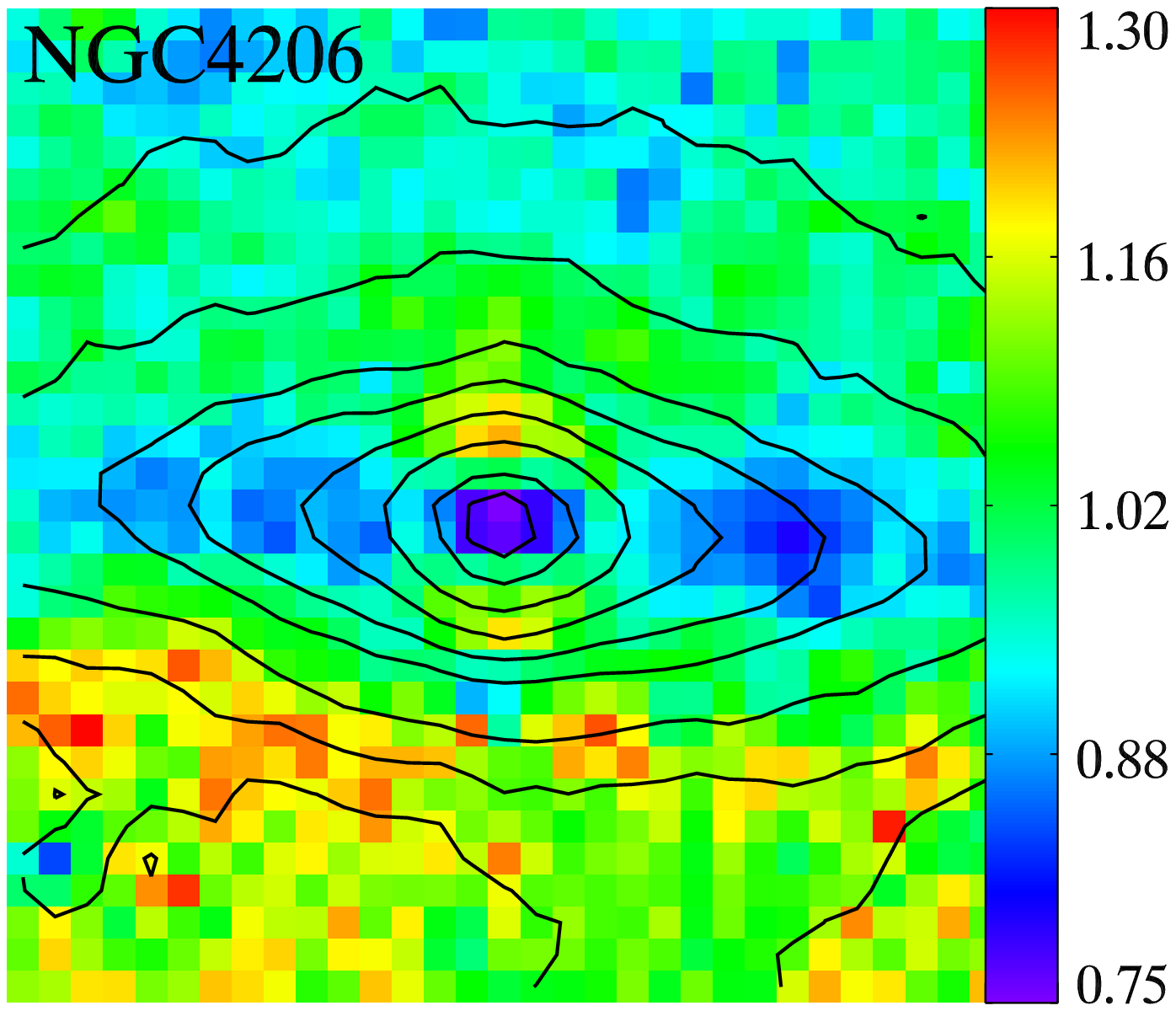}
\plotone{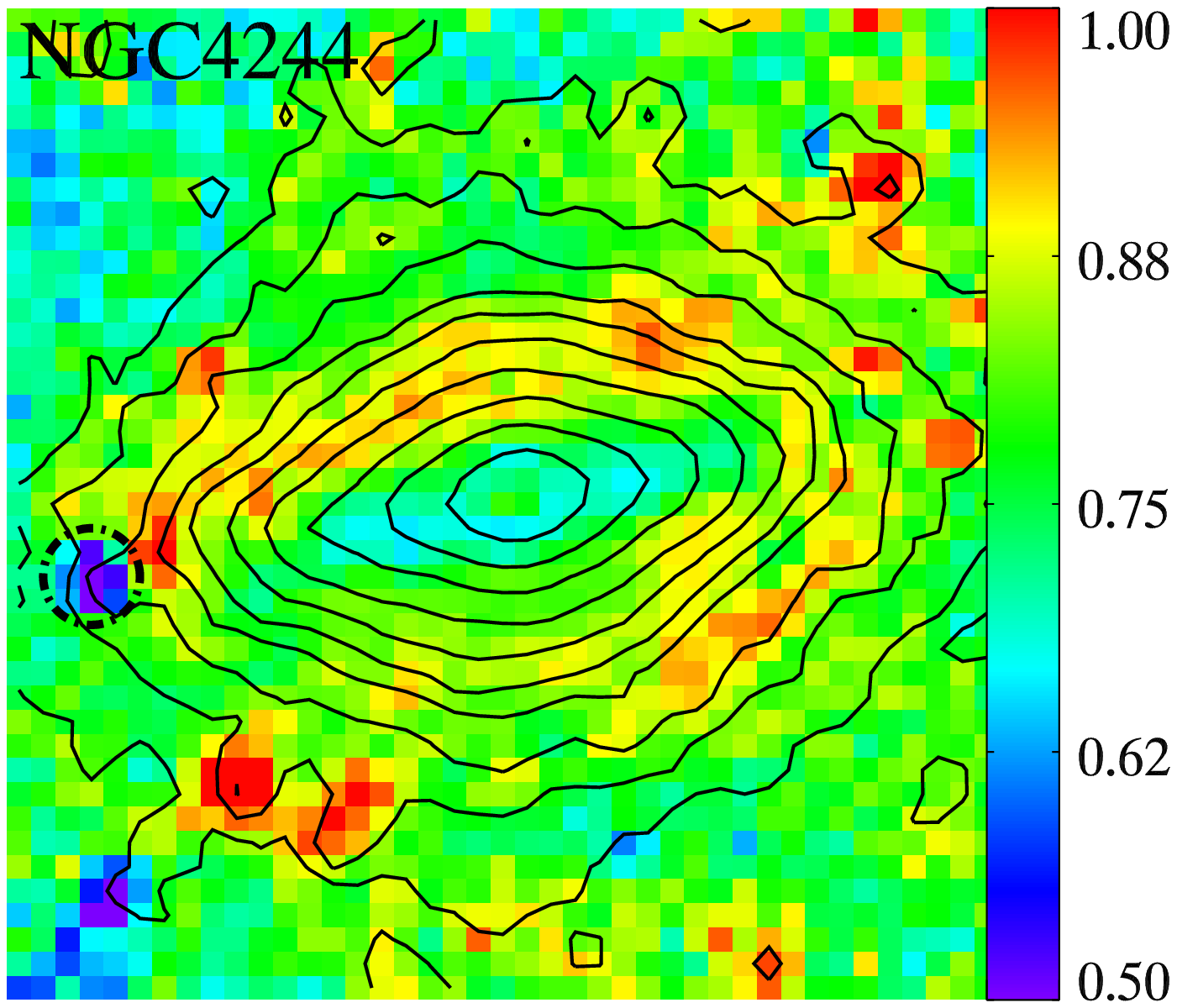}
\epsscale{1.0}
\caption{Maps of the F606W-F814W [VEGAMag] color for each cluster.
  Contours showing the F606W brightness are overlaid in black.  The
  clusters with obvious disk components (IC~5052, NGC~4206, \&
  NGC~4244) all show bluer disks and redder spheroids.  The color bars
  on the right side indicate values for the F606W-F814W color.  Colors
  have been corrected for foreground reddening.  The central eight
  pixels of the NGC~4244~NC were saturated and thus don't give
  accurate colors. The circled region in NGC~4244 shows the suspected
  HII region used to obtain a dynamical mass estimate in
  \S\ref{masssec}. }
\label{colorfig}
\end{figure*}

In contrast, the NC in IC~5052 shows two symmetric spots in its
residuals (Fig.~\ref{residfig}) and color map (Fig.~\ref{colorfig}).
Not surprisingle, the ring model provided a better fit (20\% decrease
in reduced $\chi^2$), with an inner and outer radius
of the ring of 6.8~pc and 11.8~pc, respectively.  The whole ring falls
well within the tidal radius of the spheroidal component
($\sim$24~pc), and thus their stellar distributions appear to overlap.

Nuclear rings have previously been detected in molecular gas
\citep[e.g.][]{boker97,boker98} and UV stellar light
\citep[e.g.][]{maoz96}.  However, these studies find rings an
order of magnitude larger in size than the one we find in IC~5052.
The only previously observed structure around a nuclear cluster on a
similar scale to the IC~5052 ring and NGC~4244 disk is the compact
molecular disk detected coincident with the nuclear cluster in IC~342
\citep{schinnerer03} which has a diameter of 30-55~pc depending on the
assumed distance to the galaxy.  We note that this disk is barely
resolved by their CO observations (with a resolution of 1.2\arcsec),
and may therefore be either a disk or a ring.

The evidence in all three clusters that the disk or ring components do
overlap with the spheroidal components justifies our labeling these
three nuclear clusters as multi-component clusters.  For simplicity,
and because disk models are prefered in two of the three
multi-component clusters, we will refer to the flattened components in
the next few sections as being disks.  We will return to the formation
and possible role of rings in our discussion (\S\ref{dissec}.1.3).

\subsection{Luminosities}

The F814W absolute magnitude ($M_{F814W}$) and the F606W-F814W color
are given in Table~\ref{sampletab}.  Photometry was done using the
Vega-based zeropoints and reddening coefficients from
\citet{sirianni05}. The foreground reddenings were corrected using
values of $E(B-V)$ from \citet{schlegel98}, which were at most 0.05
mags.  These measurements are complicated by the high and variable
background level and the complex shapes of the nuclear clusters.  A
scatter of $\sim$0.1 magnitudes was found by varying the aperture size
within reasonable limits.  The formal errors on the calculated
magnitudes are quite small (typically $<$0.01 mags).  Magnitudes from
the fitted models agree with the aperture magnitudes in cases where
the data is well-fit, but differ substantially when the fits are poor.

Figure~\ref{magfig} plots the F814W aperture magnitudes measured for
all nine nuclear candidates against their fitted half-light radius
from the elliptical King profiles.  We also plot magnitudes and sizes
of other compact stellar systems, assuming that $M_I \sim M_{F814W}$
since the F814W filter is very similar to the Johnson $I$ filter, with
single-stellar populations expected to be $\sim$0.03 magnitudes
brighter in F814W than in $I$ \citep{girardi06}.

Examination of Figure~\ref{magfig} shows that the magnitudes and sizes
of the objects in our sample are similar to the NCs found by
\citet{boker02,boker04a} ({\it filled circles}) for their sample of
face-on, late-type galaxies.  \citet{boker04a} measure sizes for only
the brighter clusters in their sample, but find NCs with measured
magnitudes as faint as $M_I = -8.6$.  Because of their galaxies'
face-on orientations, clusters with disk and spheroid components such
as we have observed would have gone unnoticed in their study.  The
morphology of the M33 nuclear cluster is flattened ($\epsilon = 0.17$)
along an axis parallel to the major axis of the galaxy
\citep{lauer98,matthews99}, as would be expected for a multi-component
cluster observed at the inclination of M33's disk.  The stellar
populations in the spectra of both M33's NC \citep{long02} and some of
the objects in the \citet{boker02} sample \citep{walcher06}, are also
similar to what we see in \S\ref{specpopsec} for NGC~4244's NC.
It is therefore clear that the nuclear clusters observed here are
similar to those found previously.  Furthermore, the multi-component
morphology may be a common feature of nuclear clusters.

We also note that the two most luminous clusters are those with the
most prominent disk components (NGC~4206 \& NGC~4244).  This
correlation is not surprising, as the disks seem to be composed of
recently formed stars (see \S\ref{popsec}) and would therefore have
low mass-to-light ratios.  Over time, as the young stellar populations
age the disks will become less prominent.

The observed color of the clusters (see Table~\ref{sampletab}) can
constrain the total dust column to the center of the host galaxies.
We estimated rough lower and upper limits on the total reddening by
comparing the colors of the NCs to that of an old and red, or young and
blue stellar population.  For most of the galaxies, the F606W-F814W NC
colors of $\sim$1 suggest that the reddening lies between E(B-V) of 0
and 1.  However, for NGC~3501 \& NGC~4517 the reddening must be greater
than 0.7 mags.

\section{Nuclear Cluster Stellar Populations} \label{popsec}

\begin{figure*}
\plotone{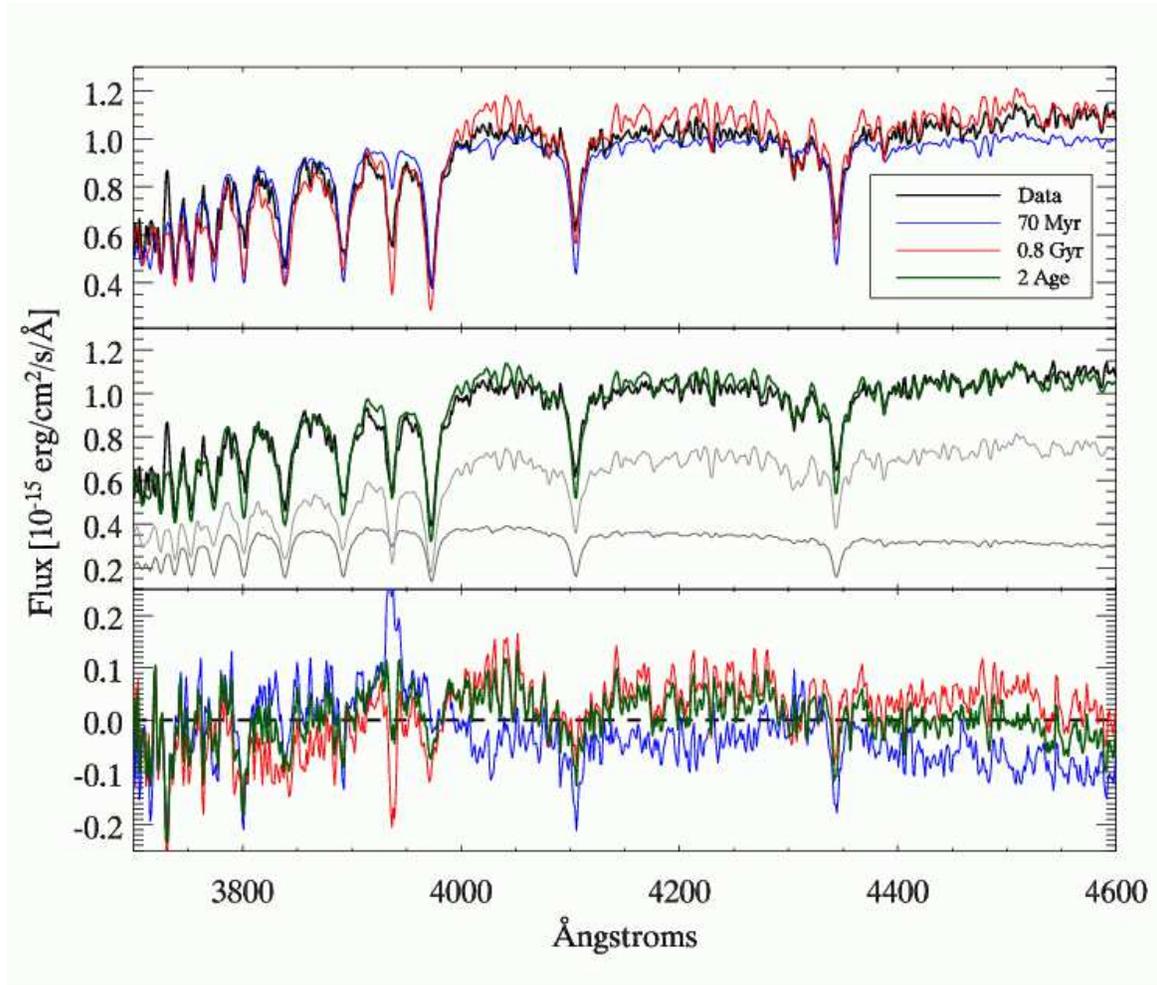}
\caption{A portion of the APO/DIS spectrum of the NGC~4244 NC with
best-fitting models overplotted.  Only the portion of the spectrum
with the most significant spectral features are shown.  {\it Top Panel
--} single-age best fits at 70 Myr and 0.8 Gyr - the ages at which
there are minima in the joint-$\chi^2$ distribution (see
Fig.~\ref{fitredfig}).  The most age-sensitive portion of the spectrum
lies around the Ca H line at $\sim$3940\AA, although numerous other
features also cannot be fit well with a single-age population.  {\it
Middle Panel --} the best two-age fit, including both stars with ages of 0.1 Gyr (dark gray line) and 1.0~Gyr (light gray line).  This fit is a significant improvement relative to either
single-age fit.  {\it Bottom Panel --} Residuals for all three fits.}
 \label{specfig}
\end{figure*}

In this section we demonstrate that the multiple structural components
seen in some NCs in \S\ref{morphsec} also have distinct stellar
populations.  In particular, the disk components are made of younger
stars than the spheroidal components.  We first examine color maps of
the clusters and argue that the disks most likely have populations
younger than 1~Gyr.  We then examine the bright nearby NC in NGC~4244
in detail, using multi-band photometry and spectroscopy (see
Fig.~\ref{specfig}) to show that the spectra is best fit by two or
more epochs of star formation.

Two recent papers have presented very thorough studies of the stellar
populations of nuclear clusters in both early and late type spiral
galaxies \citep{walcher06,rossa06}.  Both use fitting of spectra to
convincingly show that nuclear clusters contain multiple generations
of stars.  For a sample of nine bright nuclear clusters in late-type
galaxies, \citet{walcher06} has shown that all of the clusters
observed had stellar populations younger than 100 Myr.  The typical
mass of the youngest stars was a few$\times10^5$~M$_\odot$.  However,
in most of the clusters, the bulk of the stellar mass had an age of
$\gtrsim$2.5~Gyr.  Our findings on the stellar populations in our
multi-component clusters, presented below, are fully consistent with
these papers.

\subsection{Color Maps} \label{colorsec}

Figure~\ref{colorfig} shows F606W-F814W color maps for the three nuclear
cluster candidates identified as having multiple components.  In each
cluster, an obvious and significant color difference exists between
the disk and spheroidal components.  The disks are 0.3-0.6 magnitudes
bluer in F606W-F814W than the reddest parts of the spheroids.  

The observed color difference between the disk and spheroid components
is most simply interpreted as a difference in stellar population.
Despite the unknown reddening, we can place upper limits on the age of
the disk component.  Using the Padova single stellar population models
\citep{girardi06}, for a 10~Gyr old stellar population
with metallicities between [Fe/H] of -1.7 and -0.4, the F606W-F814W color
ranges from 0.73 to 0.91, with redder colors for more metal-rich
populations.  The difference in color between a 10~Gyr and 1~Gyr
population is $\lesssim$0.25 magnitudes regardless of metallicity,
suggesting that disks cannot be composed solely of intermediate age
($>$1~Gyr) stars.  At ages younger than 1~Gyr, however, the color
quickly becomes bluer.  Thus the $>0.3$~mag color difference is
evidence that the disks have stellar populations younger than 1~Gyr.
Put another way, we find that the disk components all must have
intrinsic colors bluer than F606W-F814W=0.65, which corresponds to
single-stellar populations younger than 1~Gyr.  Our analysis of the
spectrum of NGC~4244's NC (\S\ref{specpopsec}) confirms that stars with
ages $\lesssim$100~Myr are present.  Furthermore, each of the nine
nuclear cluster spectra analyzed by \citet{walcher06} showed evidence
for stellar populations with ages $<$100~Myr
 
The suggested interpretation of the color difference as a difference
in stellar population may be complicated by a number of factors.
First, even if two separate components exist, both likely contribute
to the observed color at most locations.  This overlap suggests that
the actual spread in age is larger than implied by the color
differences.  Second, the colors could be affected by the presence of
dust structures on the same scale as the clusters or by emission
line contribution.  Based on the symmetry of the nuclear clusters
(e.g. the disk is roughly the same color on both sides of the
nucleus), and the spectral results from \S\ref{specpopsec} we believe
that the color difference is unlikely to be due solely to dust or
emission lines and instead really does reflect a difference in
stellar population.

\subsection{Stellar Populations in NGC~4244} \label{specpopsec}

In the previous section we showed that the color offset between the
disk and spheroid components of the nuclear cluster indicates that the
disk has a systematically younger stellar population.  However because
of the uncertain reddening, the ACS colors alone cannot accurately
constrain the ages of the stellar populations.  In this section we use
our APO spectrum and photometry of the NC in NGC~4244 from the Sloan
Digital Sky Survey (SDSS) and 2MASS survey to constrain the cluster's
age.  Before describing the details of this analysis we first describe
the photometry.

The nuclear cluster in NGC~4244 is prominent, and stands well above
the emission from the galaxy disk.  Because of this, it was detected
as a point source in both the SDSS and 2MASS surveys.  This data
compliments the spectroscopic data by giving a large wavelength
baseline to constrain the mass, reddening, and ages of the NC
\citep{degrijs03}.  In both surveys, the source is essentially
unresolved, and photometry is determined from PSF-fitting.  In the
SDSS survey, the DR4 object ID is 58773909806344622, with PSF
magnitudes of 17.81, 16.62, 16.18, 15.96, and 15.73 in the $u'$, $g'$,
$r'$, $i'$, and $z'$ filters.  In 2MASS, the object ID is
12172945+3748264, with magnitudes of 14.36, 13.75, and 13.55 in $J$,
$H$, and $K$ bands.  The errors on the SDSS magnitudes are $\sim$0.02
mags, while the 2MASS observations have errors of $\sim$0.05 mags.
The difference in resolution and photometric methods between the two
surveys means there may be some relative systematic error in the
magnitudes.  Therefore, any conclusions based on the photometry alone
should be considered with caution.  We note that because the central
pixels of the NC were saturated in the ACS data (taken with
gain=1~$e^-$), we were unable to derive accurate photometry from the
HST images.

\subsubsection{Single-Age Fits}

In principle, the photometry and spectrum can be fit with arbitrary
stellar populations, but in practice this leads to far too many
degeneracies.  To simplify the problem we started by assuming the
NGC~4244 cluster contains stars of a single age and then determine a
best-fitting extinction, mass and $\chi^2$ as a function of age from
both the photometric and spectroscopic data separately.  For the
photometric data, we used the single-stellar population models from the
Padova group \citep{girardi04a} using the \citet{marigo01} treatment
of the AGB.  These models have been produced for both the SDSS and
2MASS photometric systems and assume a \citet{kroupa01} initial mass
function (IMF) between 0.1 and 100~M$_\odot$.  The extinction was
calculated in $A_v$ assuming $R_{v}=3.1$ using coefficients from
\citet{girardi04a} and \citet{cardelli89} for the Sloan and 2MASS data
respectively.  For the spectroscopic fitting, single stellar
population spectral templates from \citet{bruzual03} were used.  These
templates assume a \citet{chabrier03} IMF from 0.1 to 100~M$_\odot$.
They have 1\AA~resolution, and were redshifted and interpolated to the
wavelengths of our observed spectrum.  The reddening was applied to
the spectrum using the \citet{cardelli89} prescription.  We fit the
feature rich blue side of the DIS spectrum extending from 3650\AA\ to
5500\AA\ which contains Balmer, He~I, Mg, and Ca~H \& K lines (see
Fig.~\ref{specfig}).  The signal-to-noise of this portion of the
spectrum ranged from 20 to 60.  For both the photometry and
spectroscopy, we determined the best-fitting mass and reddening at a
given age using the IDL {\tt CURVEFIT} routine.  All fits were done
assuming a distance modulus of 28.20 \citep[D=4.4~Mpc,][]{seth05a},
and a metallicity of [Fe/H]=-0.4, similar to the expected current gas
phase metallicity of NGC~4244 \citep{garnett02}.  
This is the same metallicity derived from composite age fits for a
majority of the clusters found by \citet{walcher06}.  Furthermore,
\citet{rossa06} and \citet{walcher06} have shown that the inferred age
distribution does not depend strongly on the metallicity.

\begin{figure}
\plotone{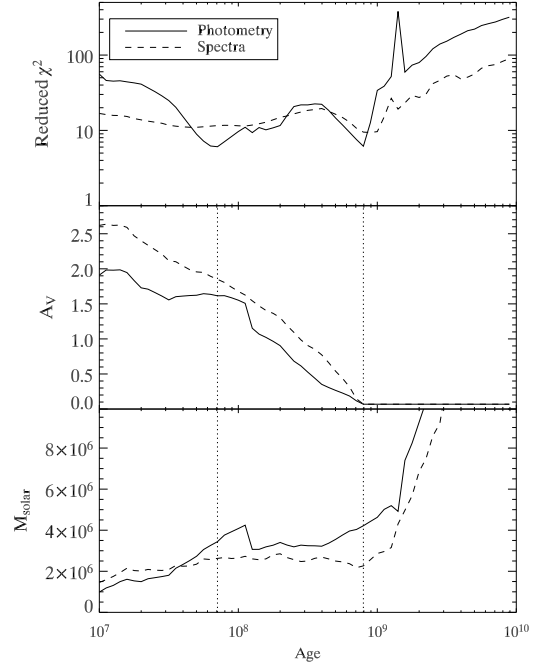}
\caption{Photometric and spectroscopic fits of the mass and reddening
of the NGC~4244 NC using single-stellar population models.  {\it Top
panel --} the reduced $\chi^2$ values of the best-fitting mass and
reddening at each age.  {\it Middle panel --} The best-fitting
reddening, {\it bottom panel --} the best fitting mass.  The dashed
vertical lines indicate the two minima in the joint $\chi^2$
distribution at 70~Myr and 0.8~Gyr.}
 \label{fitredfig}
\end{figure}

Figure~\ref{fitredfig} shows the results of our fits for single
stellar populations with ages between 10$^7$ and 10$^{10}$ years for
both the spectroscopy and photometry.  The top panel indicates the
$\chi^2$ of the best-fitting model at each age.  The spectrum is best
fit by populations younger than $\lesssim$10$^9$~years, stellar masses
of 2-3$\times$10$^6$~M$_\odot$ and decreasing extinction with
increasing age.  The $\chi^2$ has a minimum at ages of
$\sim$0.8-1.0$\times$10$^9$~years (just before the onset of the RGB),
and then increases rapidly towards older ages.  This results from the
template spectrum becoming redder than the observed spectrum.  
A shallower $\chi^2$ minima is also seen in the spectra at $\sim$50-100
Myr.

The fits to the photometric data have two comparable $\chi^2$ minima
at $\sim$10$^8$ and $\sim$10$^9$ years, with the younger age being
somewhat preferred.  The minima at these two ages is more pronounced
in the photometric data than in the spectroscopic data, perhaps as a
result of the photometric data's longer wavelength coverage. 
Combining the photometric and spectroscopic results, the minima in the
joint $\chi^2$ are at 70~Myr and 0.8~Gyr.  
The masses from the
photometric data at these minima are 3-5$\times$10$^6$~M$_\odot$,
somewhat more massive than the spectral data.  This difference may
result from light lost outside the 1.5\arcsec\ wide slit, uncertainty
in the absolute calibration of the spectroscopic data, and/or
differences in the assumed IMF.

Spectral fits from the two joint-$\chi^2$ minima are shown in
Figure~\ref{specfig}.  The younger age appears to fit the bluer part of
the spectrum better, while the older age more accurately fits many of
the lines and the redder continuum.  The Ca~H line is particularly age
sensitive, being much too shallow in the younger age and too deep in
the older age. 

\subsubsection{Multi-Age Fits}

The color difference between the disk and spheroid components
discussed in \S\ref{colorsec} and the recent results of
\citet{rossa06} and \citet{walcher06} strongly suggest that NCs
contain stellar populations of multiple ages rather than the
single-age populations assumed above.  We have tried to constrain the
ages and masses of the multiple stellar populations in NGC~4244's NC
in two ways: (1) by fitting a weighted average of two or three stellar
populations of different ages to the spectral and photometric data, or
(2) by fitting the spectrum to a constant star formation rate (SFR)
model from \citet{bruzual03}.  These approaches both give
significantly better fits to the data than single-stellar population
models, reducing $\chi^2$ values by 50\% or more.  However, it is also
clear that there is no unique solution; {\it the data can be fit
equally well by many different combinations of stellar populations and
reddening values.}  Note that we assumed a single extinction ($A_V$)
value for all components.

Of the nine NC spectra presented in \citet{walcher06}, our spectrum of
the NGC~4244 NC most closely resembles their spectrum of the NGC~7793
NC.  From their fits to composite stellar populations, they find that
the NGC~7793 cluster has an average stellar age of $\sim$2~Gyr, but
has young stars ($\lesssim$100~Myr) that make up 10\% of the mass.
This is qualitatively similar to what we find below for the NGC~4244 NC.  

{\it Two Age Models: } The simplest stellar population model
consistent with the color maps is a two-age stellar population: a
young population associated with the disk, and an older population
associated with the spheroid.  Taking the two minima seen in the
single-stellar population fits at ages of 70~Myr and 0.8 Gyr as a
starting point, we explored what combination of two populations gives
the best fit to both the spectroscopic and photometric data.  For a
young component with ages between 50-100 Myr and an intermediate-age
component with ages between 0.6 and 1.0 Gyr, the best fits to the
spectroscopy have $A_{V} \sim 0.5$, masses of
0.5-3$\times$10$^5$~M$_\odot$ for the young stellar population, total
masses of 2-4$\times$10$^6$~M$_\odot$, and reduced $\chi^2$ values of
$\sim$6.  However, for the photometry, the best fits favor an $A_{V} >
1$ and masses dominated by the young component.  These
photometric fits clearly do not reproduce the shape or lines in the
spectra, thus we restricted ourselves to lower $A_{V}$ solutions.  When
$A_{V}$ is restricted to be less than one, the photometric fits are in
relatively good agreement with the spectroscopic fits.  However, the
implied stellar masses for both the young and intermediate-age
components are up to two times larger than the spectroscopic best-fit
models, similar to what was seen in the single-age fits.  Overall, we
find that ages of 0.1 and 1.0 Gyr do the best job of matching both
sets of data.  The spectroscopic best fit at these ages is shown in
Figure~\ref{specfig}, and has $A_{V} = 0.46$,
1.9$\times$10$^5$~M$_\odot$ of 0.1 Gyr old stars and
3.1$\times$10$^6$~M$_\odot$ of 1 Gyr old stars.

Based on the color map of the nuclear cluster, we expect the younger
stellar component in the two-age fits be associated with the disk.  We
can test the consistency of this idea by comparing our spectral fits
to the luminosities of the disk component decomposed from the HST data
in \S\ref{morphsec}.  Assuming that the disk is dominated by
$\sim$70~Myr old stars, the expected F606W-F814W color from the Padova
models is 0.25 \citep{girardi06}.  The observed disk component color
is 0.4, requiring an extinction $A_V=0.47$~mags.  Using this
extinction and mass-to-light ratios from the Padova models, we derive
a mass for the disk component of 1.2$\times$10$^5$~M$_\odot$.  {\it
Both the mass and extinction derived from the luminosities of the disk
component in the HST data agree with the young stellar mass and
extinction determined in the two-age spectral fits.}  This agreement
strongly suggests a link between the blue disk component and the young
stellar component that appears to be required to fit the blue spectral
features.

{\it Three Age Models: } We also explored the sensitivity of our fits
to the inclusion of a third, old (10~Gyr) stellar population.  We find
that a large mass of old stars is consistent with both the photometric
and spectroscopic data.  Best-fits to both data sets have low $A_V$
($<$0.4), total masses of $\sim$10$^7$~M$_\odot$ with 90\% of that
mass in the old component, and $\sim$10$^5$~M$_\odot$ in the young
component, similar to the two-age fits.  The inclusion of the older
component significantly improves the photometric fit to the data
(reduced $\chi^2 = 1.7$), suggesting older stars may indeed be
present.  However, the spectral fit is insensitive to this old
component; the best-fit spectrum is nearly identical to the two-age
fits, despite 90\% of the mass being contained in the old component.

The effects of disk contamination on the photometry and spectra should
be minimal given the prominence of the NGC~4244~NC.  Futhermore, by
comparing HST/STIS and ground-based spectra, \citet{walcher06} show
that the stellar population immediately surrounding the nuclear
cluster in late-type galaxies seems to have a population similar to
the cluster itself.  Both \citet{rossa06} and \citet{walcher06} find
some evidence for an underlying old stellar component in many of their
nuclear cluster spectra.  We note that all the masses quoted here
refer to the {\it initial} mass in stars required to fit the current
spectral data.  Due to stellar evolution, the current mass of a
cluster dominated by old stars would be $\sim$2$\times$ lower than the
initial mass, given reasonable initial mass functions.

{\it Constant SFR model: } Using the \citet{bruzual03} codes, we
generated a model spectrum assuming a constant SFR from 12~Gyr to the
present.  The best-fit of this model to the data gives a current mass
of the cluster of 6.2$\times$10$^6$~M$_\odot$, and a reddening of
$A_{V} = 0.72$. 
The reduced $\chi^2$ of the fit is 5.7, within 10\% of
the reduced $\chi^2$ of the best fits obtained for the two- and
three-age models.  The spectral shape of the constant SFR model
appears to be somewhat redder than the data, suggesting either the
need for relatively more young stars than in the constant SFR model,
or for a more realistic metallicity evolution, in which the older
stars are bluer and more metal-poor.  The former point is supported by
\citet{walcher06}, who have shown that in their sample of clusters,
the star formation rate in the past 100~Myr is an order of magnitude
greater than the star formation rate between 0.3 and 20~Gyr.  However,
this enhancement of recent star formation may result from a bias in
their sample towards brighter nuclear clusters.

\subsubsection{Summary of Stellar Populations in the NGC~4244 NC}

We draw the following conclusions from the spectral and photometric
fits of the NGC~4244~NC:
\begin{enumerate}
\item Multiple stellar populations definitely provide improved fits
over single-stellar populations for both the spectroscopy and
photometry of the NGC~4244 NC.  However, many different combinations
of masses and ages fit the data well.  
\item The best-fitting single stellar populations have ages of either
$\sim$70~Myr or $\sim$0.8~Gyr.
\item The total current mass of the cluster
appears to be between 0.2 and 1$\times$10$^7$M~$_\odot$.
\item If we assume only two ages are present, the youngest ($\lesssim$ 100
Myr) stellar component appears to have $\sim$10$^5$~M$_\odot$ of
stars.  This mass is consistent with the luminosity of the disk
component decomposed using the morphological fits in \S\ref{morphsec}.
\item For all multi-age fits, the mass of stars younger than 100~Myr
is a small fraction ($<5$\%) of the total mass of the cluster.
\item Stars older than 1 Gyr may be present in large numbers
($\sim$10$^7$~M$_\odot)$.  Inclusion of an old stellar population
particularly improves our fits to the SDSS and 2MASS photometric data.
\item A constant SFR fits the data as well as assuming two or three
discrete ages.
\end{enumerate}

\section{A Dynamical Mass Estimate for the NGC~4244 cluster} \label{masssec}

\begin{figure}
\plotone{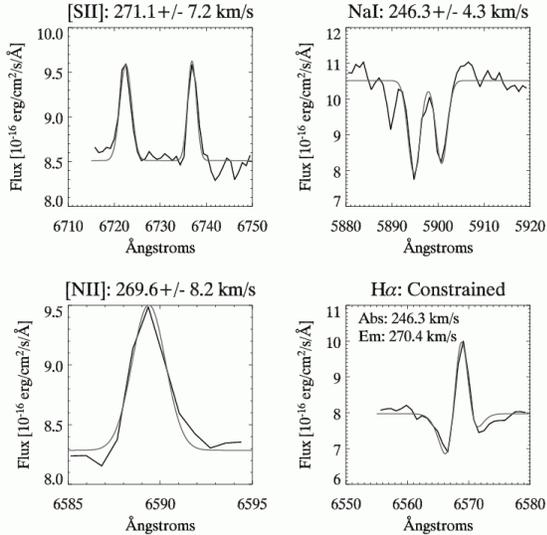}
\caption{Spectral lines used to derive the relative offset between
emission and absorption.  Dark lines show the actual spectrum, while
lighter lines show the best fitting gaussian or double-gaussian
model. The number in the title of each panel gives the fitted velocity
and error bar. The spectra has a S/N$\sim$35/pixel.  }
 \label{findmassfig}
\end{figure}

Emission lines, including H$\alpha$, [NII], [SII], and [OII] are
seen in our spectrum of the NGC~4244 NC.  For the H$\alpha$ line, the
emission component is redshifted relative to the underlying stellar
absorption lines (see Fig.~\ref{findmassfig}).  We utilize this offset
between the emission and absorption components to obtain a dynamical
mass estimate for the NGC~4244 cluster.

Examination of the color map for NGC~4244 in Figure~\ref{colorfig}
shows a very blue compact source located along the NC plane on the
left (NE) side, 0.88\arcsec (18.6~pc projected) from the NC center.
This compact blue object has an F606W-F814W color of $\sim$0.3, which
is $\sim$0.4-0.6 mags bluer than the disk and spheroid components of
the cluster.  Assuming moderate reddening and blending with the outer
isophotes of the cluster, the color of the blue knot is consistent
with this object being an HII region.  HII regions in our filter
combination appear very blue because there are strong emission lines
located in the F606W filter and not in the F814W filter.  Measurements
of HII regions in NGC~55 (the nearest galaxy in our sample) suggests
HII regions have typical colors of F606W-F814W$=$-0.2 with a $\sim$0.2
mag scatter.

This likely HII region falls within the slit used for our nuclear
cluster spectrum.  Therefore, we identified this object as the likely
source of the emission lines in our spectra.  We verified this by
extracting spectra from the left (NE) and right (SW) sides of the
cluster.  The left side spectrum shows significantly enhanced emission
lines relative to the right side.  However, because the seeing was
$\sim$1\arcsec, we were unable to spatially separate the emission
component from the nuclear cluster.  We now proceed under the
assumption that the emission lines in our spectrum are associated with
this compact blue object.

To determine the relative velocity of the emission and absorption
components we fit the [SII] doublet and [NII] line to determine the
emission line velocity, and fit the NaI doublet to determine the
absorption line velocity.  
We obtained a weighted mean velocity of 270.5$\pm$5.4 km/sec for the
emission lines and 246.3$\pm$4.3 km/sec for the NaI absorption line
assuming air wavelengths for the lines from the Atomic Line
List\footnote{http://www.pa.uky.edu/\~{}peter/atomic/}.  This
absorption velocity matches the systemic velocity of NGC~4244
\citep[244.4$\pm$0.3~km/sec,][]{olling96} to within the error bars, as
would be expected for the nuclear cluster.  We note that this HII region
is counter-rotating with respect to the galaxy disk. 

Figure~\ref{findmassfig} shows the fitted emission and absorption
lines, as well as the best fit of absorption $+$ emission for the
H$\alpha$ line after fixing the velocities of the absorption and
emission to the values derived from the NaI, [SII], and [NII] lines.  This
H$\alpha$ line fit does a good job of replicating the observed line
shape thereby suggesting that the velocities we have derived are
correct.  

Assuming the projected radius of the HII region is close to the true
radius, and that it is bound to the NGC~4244 NC nuclear cluster, the
$\Delta v = 24.1 \pm 7.0$ km/sec offset between emission and absorption
implies a nuclear cluster mass of 2.5$^{+1.7}_{-1.2} \times
10^{6}$M$_\odot$ within a radius of 19~pc.   If the object is
indeed rotating around the cluster, but is in front of or behind the
cluster (thus making its projected distance larger), then the mass we
derived is a lower limit on the mass of the cluster.

This rough estimate of the dynamical mass of NGC~4244's NC is
consistent with the stellar masses determined for the one- and two-age
fits in \S\ref{specpopsec}.  It is also consistent
with the constant SFR or three-age model fits
(M$\sim$6$\times$10$^6$~M$_\odot$) if the HII region is seen in
projection and lies $\sim$25~pc from the center of the nuclear
cluster.

The estimated mass is also similar to the typical mass determined by
\citet{walcher05} for late-type galaxy nuclei.  They derived masses
for nine nuclear clusters from the \citet{boker02} sample by combining
size and velocity dispersion measurements.  Their total range of
masses is 8$\times$10$^5$~M$_\odot$ to 6$\times$10$^7$~M$_\odot$, with
seven of the nine clusters having masses between 0.1 and
1$\times10^7$~M$_\odot$.  We note that they have determined these
masses assuming spherical symmetry; if these clusters have disk
components similar to the ones we have observed here, the masses
derived by \citet{walcher05} may be underestimates, since they do not
take into account the rotational support that would be expected in a
disk.  Given that their clusters contain 
young stellar populations \citep{walcher06}, this effect could be significant.

\section{NGC~4206: Possible Indication of an AGN Component} \label{sbhsec}

In this section, we present evidence that the nuclear cluster in NGC~4206 may
contain an AGN component.  This evidence is very tentative and requires
confirmation with follow-up observations.

An SDSS spectrum (MJD 53149, Plate 1612, Fiber 603) exists for the
central regions of NGC~4206.  Due to the 3\arcsec\ diameter
($\sim$250~pc) aperture, this spectrum is dominated by the light from
the central regions of the galaxy and not by the nuclear cluster.  In
addition to a dominant old stellar spectrum, narrow H$\alpha$, [NII],
and [SII] emission are clearly seen.  No broad emission lines are
seen.  Examining the HST color map, the
center of the nuclear cluster is the bluest area anywhere within the
SDSS spectral aperture.  Since emission line sources are very blue in
our filter combination, the central portion of the nuclear cluster is
the likely source of the emission lines.  The emission lines are not
offset in velocity from the absorption spectrum (as was seen in
NGC~4244) further supporting their nuclear origin. 

The ionizing source of the observed emission lines could be either
star formation or AGN activity at the center of the nuclear cluster.
These two ionizing sources can be distinguished using the line ratio
of [NII] and H$\alpha$.  Because the H$\alpha$ line is contaminated by
stellar absorption, we model the underlying stellar continuum with a
single stellar population $\sim$3~Gyr in age with $A_V \sim 0.5$.
After subtracting this spectrum, the line ratio of the remaining
emission lines is log([NII]/H$\alpha$)$ \sim -0.3$.  Based on the
recent analysis of SDSS spectra by \citep{obric06}, this line ratio
suggests that the source is a ``mixed'' starburst/AGN.  NGC~4206 may
therefore host a small AGN within the core of the nuclear star
cluster.  If confirmed by follow-up observations, this would make it
only the second late-type spiral known to have both a SMBH and massive
nuclear star cluster, the other being NGC~4395, an Sd galaxy with $M_B
\sim -17.5$ hosting a nuclear cluster with $M_I \sim -10.1$
\citep{matthews99}, and a black hole with a mass of $\sim$10$^4 -
10^5$~M$_\odot$ \citep{filippenko03}.  In earlier type spirals, other
examples exist of the co-existence of nuclear star clusters and AGN
\citep[e.g.][]{thatte97,davies06}.  We plan to test for the presence
of AGN components in all of our multi-component nuclear cluster
candidates with near-IR integral field unit spectroscopy using
adaptive optics.


\section{Discussion} \label{dissec}

We now discuss how our data constrains the formation mechanisms of
nuclear clusters.  We suggest a formation mechanism in which stars in
nuclear clusters form episodically in disks and then over time lose
angular momentum or are heated vertically, ending up in a more
spheroidal distribution.
After discussing this mechanism in some detail, we examine the
relation of nuclear clusters to other massive star clusters and
supermassive black holes.

\subsection{Formation Mechanisms}

Our observations of multi-component clusters place significant
constraints on nuclear cluster formation mechanisms.  The presence of
young NC disks ($<$1~Gyr) aligned with the disks of the host galaxies
strongly suggests that they are formed {\it in situ} from gas accreted
into the nuclear regions.  We have shown in \S\ref{morphsec}.2 that
these disks may be a common feature of nuclear clusters.  Such disks
cannot be formed via the accretion of globular clusters by dynamical
friction.  While globular cluster accretion very likely occurs, this
mechanism has been shown to be relatively inefficient in late-type
galaxies \citep{milosavljevic04}.  We propose
here a scenario in which the {\it dominant} formation mechanism for
nuclear clusters is via {\it in situ} formation in disks.
Although the mechanism will need to be tested with detailed
simulations, it is consistent not just with our data, but with both
the evidence for young and composite stellar populations in many
nuclear clusters \citep[e.g][]{walcher06}, and the direct detection of
a molecular gas disk coincident with the nuclear cluster in IC~342
\citep{schinnerer03}.  

On a large scale ($\sim$1~kpc), both dissipative merging of
molecular clouds and magnetorotational instabilities have been shown
to be capable of supplying the nucleus of a late-type galaxy with
sufficient gas to form nuclear clusters
\citep{milosavljevic04,bekki06}.  Our proposed {\it in situ} formation
mechanism is concerned with what occurs once the gas reaches the center.
Our model is also fully consistent with the suggestion by
\citet{walcher06} that an initial seed star cluster is required for
subsequent nuclear star cluster formation.

The details of an {\it in situ} formation mechanism can be constrained
by our observations.  To explain presence of redder/older spheroids in
the multi-component clusters and the absence of disk components in a
majority of our nuclear clusters, we propose a model in which nuclear
cluster disks form episodically and are transformed into a rounder
distribution over time due to loss of angular momentum or dynamical
heating.  There are two important timescales in this model: (1) the
period of time between star formation episodes, and (2) the time it
takes stars in a disk morphology to end up in a spheroidal
distribution.  We examine these two timescales in detail below.

\subsubsection{Period Between Star Formation Episodes}  

A number of authors have previously suggested that star formation in
the nuclear regions of late-type spirals may be episodic.
The physical motivation for episodic star formation is that a massive
star formation event will feedback on the local environment preventing
star formation for the next $\sim$10$^8$ years
\citep{milosavljevic04}.  Although this feedback may destroy the gas
disk and shut off star formation, it does not disrupt the disk of
recently formed stars, whose lifetime we discuss below in
\ref{dissec}.1.2.

Assuming that the present epoch is not a special time in the life of
the nuclear clusters, one can estimate the period on which star
formation recurs.  Observationally, \citet{davidge02} argue for a
period of $\sim$0.3~Gyr based on the detection frequency of Pa$\alpha$
emission in galaxy nuclei. \citet{schinnerer03} also determine a
recurrence period of a few hundred Myr to 1 Gyr based on the gas
accretion rates observed in IC~342's nucleus.  More recently, from
analysis of the youngest stellar components in the spectrum,
\citet{walcher06} has found the typical time between star formation
bursts to be $\sim$70~Myr with a typical mass of roughly
$1.6 \times 10^5$~M$_\odot$.

Our observations of the NGC~4244 nuclear cluster are consistent with
these previous estimates.  Based on the $\sim$10$^5$~M$_\odot$ of
stars in the young component (\S\ref{specpopsec}), approximately 25
episodes of disk formation must have occurred to build up the
2.5$\times$10$^6$~M$_\odot$ of the cluster (\S\ref{masssec}).  If this
occurs over a Hubble time, then the period between bursts is
$\sim$0.5~Gyr.  

\subsubsection{Timescale of Disk Transformation}

Given that nuclear cluster disks are observed in $<$50\% of nuclear
clusters, and that these disks all have stellar populations younger
than a Gyr, a mechanism must exist that destroys or disrupts this disk
on a timescale shorter than $\sim$1~Gyr.  We suggest that angular
momentum loss or dynamical heating causes the disks to be transformed
over time, thus creating the more spheroidal, older components seen in
our nuclear clusters.  Although mechanisms of angular momentum loss
have not been studied previously in nuclear clusters, these objects
are dynamically similar to globular clusters, for which the theory of
rotation and rotational evolution has been well-developed
\citep{agekian58,king61,shapiro76,fall85,davoust90,longaretti97}.

As they age, star clusters flattened due to rotation will become
increasingly round due to the preferential evaporation of high angular
momentum stars \citep{agekian58}.  We can compare the timescale of
this angular momentum loss to the expected lifetimes of our disks.
The timescale over which the ellipticity of a rotating stellar cluster
decreases is $\gtrsim$40$\times$ the median relaxation time,
$\tau_{rh}$ \citep{shapiro76,fall85}.  For globular clusters in the
Milky Way, $\tau_{rh}$ is typically $\sim$10$^8$ years.  However, this
timescale is longer for more massive clusters like $\omega$~Cen, with
$\tau_{rh} = 3.5$~Gyr \citep{shapiro76}.  For the NGC~4244 NC we have
calculated a median relaxation time for the whole cluster of
$\sim$2 Gyr \citep[Eq. 8-72][]{binney87}.  For just the disk
component, assuming it has a mass of $\sim$10$^5$~M$_\odot$ and a half
mass radius of $\sim$3 pc, we get $\tau_{rh}=4\times10^8$ years.
These relaxation times suggest that the timescale for evaporation of
angular momentum in NGC~4244 is in the range of 16-80~Gyr.  This
timescale is significantly longer than the 100s of Myrs over which we
expect the disk to be transformed into a spheroidal component.

It therefore appears that angular momentum loss due to processes
internal to the clusters are unlikely to transform the disk components
into spheroids on sufficiently short timescales.  Instead, we must
look to external processes that can affect the the dynamical evolution
of the nuclear cluster.  We 
have identified a few alternative mechanisms for transforming the
stellar disk component into a spheroid.  These include: (1) small
angular offsets of the nuclear disk from the plane of the galaxy, and
(2) slight positional offsets of the cluster from the center of the
galaxy potential, (3) torques on the stellar disk from successive gas
accretion.  
A forthcoming paper will discuss this problem in more detail.

\subsubsection{Rings}

We have been assuming thus far that stars are formed initially in
disks, as suggested by our observations of NGC~4244 and NGC~4206.  
However, in \S\ref{morphsec}.1.3, we showed that one of the three
multi-component NCs, in IC~5052, was best fit by a spheroid $+$ ring
model with inner and outer ring radii of 6.8 and 11.8~pc.  We discuss
here how such a ring could form and examine rings in context of our
proposed nuclear cluster formation model.

Based on an elongated near-infrared component in IC~342,
\citet{boker97} suggested that the observed molecular ring (with
radius\footnote{The radius of their ring was
recalculated assuming the Cepheid distance to IC~342 of 3.3~Mpc derived
by \citet{saha02}.} of $\sim$65~pc)  results from an inner Lindblad resonances (ILR)
with a stellar bar.  If a similar mechanism is at work in forming the
ring in IC~5052, it would require the star cluster itself to have a
barred disk morphology.  From HST imaging of nuclear clusters in
face-on spirals, there is no evidence for bars on these scales
\citep{boker04a}.  A ring could also result from tidal stripping of an
infalling molecular cloud or star cluster.  Assuming a pre-existing
cluster with mass of $3 \times 10^6$~M$_\odot$ and a $10^5$~M$_\odot$
molecular cloud, the molecular cloud would be tidally stripped between
radii of 8 and 20~pc for molecular cloud number densities ranging
between $5 \times 10^4$ and $5 \times 10^3$~cm$^{-3}$.  Thus, tidal
stripping of a molecular cloud is a feasible mechanism for creating a
ring such as the one in IC~5052.

Episodic star formation occuring at least occasionally in rings is
consistent with the proposed formation model.  The ring in IC~5052 is
internal to the cluster itself, so after loss of angular momentum or
dynamical heating, its stars would still contribute to the stellar
distribution of the spheroid.  The radius of the ring depends on the
current mass of the star cluster in both mechanisms for ring formation
described above.  This correlation might provide a natural explanation
for the clear correlation of nuclear cluster size and luminosity
observed in dwarf elliptical galaxies by \citet{cote06}.  A similar
explanation could work for formation in disks if the size of the disks
depends on the mass of the nuclear cluster.  Detailed modelling of
disk/ring formation and dissipation would be required to test the
feasibility of this idea.

\subsubsection{Other Formation Scenarios}

Our proposed scenario naturally explains the formation of
multi-component clusters.  However other scenarios may also be
consistent with our observations.  While our observations seem to
require some {\it in situ} formation to explain the presence of young
stellar disks, they do not rule out the possibility that the
multi-component nuclear clusters were formed in part from the merging
of globular clusters \citep[e.g.][]{tremaine75,lotz01}.  In this
``hybrid'' scenario, the young disk components may be a late-time
veneer on an already massive cluster formed from merged globular
clusters.  However, we note that \citet{milosavljevic04} has
calculated that the timescale for dynamical friction in late-type
spirals is $\gtrsim 3 \times 10^{9}$ years even for relatively massive
clusters (10$^6$~M$_\odot$) located within the central kpc of the
galaxy.  This quantity increases for less massive and more distant
clusters.  He concludes that this timescale is thus too long to be the
dominant mechanism for nuclear cluster formation, and shows that
magnetorotational instabilities can transport sufficient quantities of
gas into the galaxy center to account for the observed NCs in
late-type spirals.

\subsection{Relationship to Other Massive Star Clusters}

Our observations have implications beyond just the formation of
nuclear star clusters in late-type galaxies.  We briefly review here
the possible connections between late-type spiral nuclei and dwarf
elliptical nuclei, and the evidence that some massive globular clusters and
ultra-compact dwarfs are stripped nuclear clusters.

\subsubsection{The relation between spiral and dE nuclei}

The recent study by \citet{cote06} shows that the nuclei of dE and
late-type spiral galaxies have comparable sizes, luminosities, and
frequencies of detection (see also Fig.~\ref{magfig}).  They
argue that this similarity suggests a nuclear cluster formation
mechanism that does not depend strongly on galaxy properties.
However, the formation scenario suggested by our observations for
nuclear clusters in late-type galaxies cannot take place in the
gas-free dE galaxies.  We therefore interpret the similarity of the
nuclei in these galaxies as support for dE galaxies evolving from
low-mass spiral and irregular galaxies \citep[e.g.][]{davies88}.  In
this scenario, a dE,N galaxy would result from a nucleated late-type
spiral galaxy after depletion of gas through star formation or galaxy
harrassment \citep{moore96}.  If so, then the ages of the nuclear
clusters in dE galaxies can constrain the epoch when these systems
were stripped.

\subsubsection{Massive star clusters as stripped galaxy nuclei}

There is direct evidence that some massive globular clusters are
associated with stripped galaxies.  The galactic
globular cluster M54 appears to be the nucleus of the Sag dSph
\citep{layden00}, and six of the most massive globular clusters in
NGC~5128 appear to be surrounded by extratidal light
\citep{harris02,martini04}.

Stripped nuclear clusters have also been invoked to explain the
multiple stellar populations observed in the most massive nearby
globular clusters \citep[e.g][]{gnedin02}.  These observations include
the large metallicity spread \citep{smith04} and double main-sequence
\citep{bedin04} in $\omega$~Cen, and the spread in red giant branch
colors for G1 in Andromeda \citep{meylan01}.  More recently,
\citet{harris06} have observed a trend towards redder colors with
increasing luminosity for the brightest blue globular clusters in
massive elliptical galaxies.  This trend is interpreted as evidence
for self-enrichment in these clusters pointing to an extended star
formation history more typical of nuclear clusters than globular
clusters.  A similar redward trend is seen for dE nuclei by
\citet{cote06}.  Self-enrichment is naturally explained in an {\it in
situ} formation mechanism, but is not easily explained if these
clusters are formed by merging less massive clusters via dynamical
friction.

The ultracompact dwarf galaxies recently found in the Fornax
\citep{phillipps01,drinkwater03} and Virgo \citep{jones06} clusters
may also be related to nuclear star clusters.  One possible mechanism
for their formation is ``galaxy threshing'', in which nucleated
galaxies lose their envelopes leaving behind only their nuclei as UCDs
\citep{bekki01a}.  In the $M_I$ vs. log($r_{eff}$) diagram in
Figure~\ref{magfig}, NGC~4206's NC has sufficient size and luminosity,
even after modest age fading, to be a UCD progenitor.  The galaxy
threshing scenario has also been shown to be feasible by \citet{cote06}
for dwarf elliptical nuclei, although \citet{depropris05} suggests
that dE,N are in general more compact than UCDs.  One of the galaxy
nuclei they study does have similar properties to the UCDs, as well as
possessing spiral arms which indicate the galaxy may be an anemic
spiral.

\subsection{Connection to Black Holes}

As residents of the same neighborhoods, it is natural to suspect that
there may be some connection between nuclear clusters and supermassive
black holes (SMBHs).  The presence of SMBHs in the nuclei of $>$30
galaxies have been confirmed through dynamical measurements of their
masses \citep[reviewed by][]{ferrarese05}.  The masses of SMBHs are
known to correlate with the mass and/or velocity dispersion of the
bulge component \citep[the $M_\bullet-\sigma$
relation,][]{ferrarese00b,gebhardt00}.  Recent work on nuclear
clusters in both spiral and elliptical galaxies have shown a
remarkably similar trend \citep{rossa06,cote06,ferrarese06,wehner06}.
For a large sample of nucleated elliptical galaxies, \citet{cote06}
have shown that the nucleus accounts for 0.30$\pm$0.04\% of the total
galaxy mass.  This fraction is very similar to the fraction of galaxy
mass found in the SMBHs \citep{ferrarese06}.  For this reason,
\citet{cote06}, \citet{ferrarese06} and \citet{wehner06} suggest that
nuclear clusters and SMBHs be thought of as a single-class of {\it
Central Massive Objects}.  They suggest a transition at $\sim$10$^7$
M$_\odot$ between nuclear clusters and SMBHs.  However, the detection of
numerous lower mass SMBHs at the centers of early and late-type dwarf
galaxies \citep{barth05,greene05}, and the co-existence of nuclear star
clusters and SMBHs in NGC~4395 \citep{filippenko03} and perhaps in
NGC~4206 (\S\ref{sbhsec}), suggest that this transition may not be
abrupt.

The formation of black holes in the centers of nuclear clusters may be
related to the intermediate mass black holes (IMBHs) found in the
centers of massive non-nuclear star clusters.  A
1.7$\times$10$^4$~M$_\odot$ black hole has been reported by
\citet{gebhardt05} in the globular cluster G1 in Andromeda (which is
also a possible stripped nuclei), based on stellar dynamics.  A
10$^2$-10$^4$~M$_\odot$ black hole has also been identified in one of
M82's young massive clusters, based on X-ray observations
\citep{patruno06}.  These black holes are thought to form either by
early merging of massive young stars within the cluster, or by
dynamical interaction in the core of the clusters over long time
periods \citep[][and references therein]{miller04}.  If IMBH formation
is common in dense massive clusters, then nuclear clusters should
frequently host black holes as well.  Subsequent mergers of many IMBHs
could then form supermassive black holes (SMBHs) \citep{miller04}.  As
the mass of the SMBH increases, these mergers will likely destroy the
nuclear clusters \citep[see \S5.2.2 of ][]{cote06}, distributing their
stars into the bulge.

\section{Summary} \label{sumsec}

We have examined the nuclear regions of fourteen nearby galaxies using
HST/ACS images in the F606W and F814W filters.  We find nuclear
cluster candidates within 2\arcsec\ of the photocenter of nine of these
galaxies.  We have also obtained multi-band photometry and a
spectrum of the NGC~4244 nuclear cluster, the nearest object in our
sample.  Analyzing these nuclear clusters, we find the following:

\begin{enumerate}
\item From the edge-on perspective, three of the nine cluster
  candidates (IC~5052, NGC~4206, NGC~4244) have morphologies
  suggesting they have both disk and spheroid components.  These
  components are very compact with scale lengths and half-light radii
  of 2-30~pc.  Star clusters with multiple visible components have not
  previously been seen.  
\item The three multi-component nuclear clusters have elongations
  oriented to within $\sim$10$^\circ$ of the major-axis of the
  galaxy.  
\item The magnitudes and sizes of the clusters in our edge-on sample
  are very similar to those observed in other face-on, late-type
  galaxies \citep{boker04a} and dE galaxies \citep{cote06}.
\item The spheroids of the multi-component nuclear clusters are
  0.3-0.6 magnitudes redder than the disks in F606W-F814W.  This color
  difference implies the disks are populated with stars $<$1~Gyr in
  age.
\item Fitting single-stellar population models to the spectral and
  photometric data of the NGC~4244 nuclear cluster, we find it is
  best-fit by ages of $\sim$70~Myr or $\sim$0.8~Gyr.  The implied total mass
  of the cluster for is $\sim$3$\times$10$^6$~M$_\odot$.
\item A combination of two or more stellar populations significantly
  improves the fit to the NGC~4244 nuclear cluster spectral and
  photometric data.  In the multi-age fits, the youngest
  ($\lesssim100$~Myr) stellar population matches the luminosity of the
  disk component, and appears to make up $\lesssim$5\% of the total
  mass of the cluster.  As much as $\sim$10$^7$~M$_\odot$ of old
  (10~Gyr) stars may also be present in the cluster.  A constant star
  formation rate over the last 12~Gyr also provides a surprisingly
  good fit.
\item Using an HII region 19~pc from the center of the NGC~4244 NC, we
  measure the velocity offset of emission and absorption components in
  our spectrum to obtain a lower limit to the dynamical mass for the
  nuclear cluster of 2.5$^{+1.7}_{-1.2} \times 10^{6}$M$_\odot$.
\item Analysis of emission lines in the NGC~4206
  nucleus suggest that it may host an AGN as well as a stellar
  cluster.
\end{enumerate}

These observational results strongly support an {\it in situ}
formation model for nuclear star clusters.  We suggest a specific
model in which stars are formed episodically when gas accretes into a
compact nuclear disk.  After formation the stars gradually lose
angular momentum and eventually end up in a spheroidal component.  We
show that a duty cycle for star formation of $\sim$0.5~Gyr is
consistent with our and previous observations.  The angular momentum
must also be lost on this timescale, which we suggest may occur due to
interactions of the nuclear cluster with the surrounding galaxy
potential.

Acknowledgments: The authors would like to thank their anonymous
referee, Carl-Jakob Walcher, Thomas Puzia, Tom Quinn, and Nate Bastian
for their helpful comments on improving this manuscript.  We also
thank Leo Girardi for his help in interpreting models, Andrew West for
his help with SDSS, Kevin Covey for fruitful discussions, and Roelof
de Jong for his ongoing support.  This work has been supported by
HST-AR-10309, HST-GO-9765, the Sloan Foundation, and NSF grant CAREER
AST 02-38683.

\end{document}